# Visible and Near-Infrared Reflectance of Hyperfine and Hyperporous Particulate Surfaces

R. Sultana[a*], O. Poch[a], P. Beck[a,b], B. Schmitt[a] and E. Quirico[a]

[a] Université Grenoble Alpes, CNRS, Institut de Planétologie et d'Astrophysique de Grenoble, Grenoble, France,

[b] Institut Universitaire de France, Paris, France

[*] Corresponding author: robin.sultana@univ-grenoble-alpes.fr

## Abstract

The composition of Solar System surfaces can be inferred through reflectance and emission spectroscopy, by comparing these observations to laboratory measurements and radiative transfer models. While several populations of objects appear to be covered by sub-micrometre sized particles (D < 1 µm) (referred to as hyperfine), there are limited studies on reflectance and emission of particulate surfaces composed of particles smaller than the visible and infrared wavelengths. We have undertaken an effort to determine the reflectance of hyperfine particulate surfaces in conjunction with high-porosity, in order to simulate the physical state of cometary surfaces and their related asteroids (P- and D-types). In this work, we present a technique developed to produce hyperfine particles of astrophysical relevant materials (silicates, sulphides, macromolecular organics). This technique is used to prepare hyperfine powders that were measured in reflectance in the 0.4-2.6 µm range. These powders were then included in water ice particles, sublimated under vacuum, in order to produce a hyperporous sample of hyperfine material (refers as to sublimation residue). When grinded below one micrometre, the four materials studied (olivine, smectite, pyroxene and amorphous silica), show strong decrease of their absorption features together with a blueing of the spectra. This "small grain degeneracy" implies that surfaces covered by hyperfine grains should show only shallow absorption features if any (in the case of moderately absorbing particles as studied here). These two effects, decrease of band depth and spectral blueing, appear magnified when the grains are incorporated in the hyperporous residue. We interpret the distinct behaviour between hyperporous and more compact surfaces by the distancing of individual grains and a decrease in the size of the elemental scatterers. This work implies that hyperfine grains are unabundant at the surfaces of S- or V-type asteroids, and that the blue nature of B-type may be related to a physical effect rather than a compositional effect.

# 1.Introduction

The Solar System hosts a population of small objects (D < 1000 km) of which only an extreme minority has been visited by spacecrafts. These objects, referred to as the small bodies population, are diverse in terms of composition and geological history. While some bodies appear to have experienced heating or full differentiation (for instance 4-Vesta and its vestoids (Drake, 2001), others appear to be more primitive in nature (Vernazza and Beck, 2016)). Comets, being rich in volatiles and "super-volatiles", are usually considered as the most primitive, meaning that they have been the least processed since their formation in early stages of the Solar System. Small bodies can be classified according to their mechanical dynamic (i.e. non-exhaustively Near-Earth-Asteroids, Main-Belt Asteroids, Jupiter Trojans, long and short-period comets) or to their surface optical properties. Surface mineralogy can be estimated by means of ground-based observation of reflected sunlight spectra (for the largest objects only). Because only some of the compounds making up the surface can produce absorption in the spectral range covered by these passive reflectance spectra, the determined mineralogy is strongly biased.

In order to have a more robust estimate of the composition a widely used approach has been to compare reflectance spectra of meteorites (extra-terrestrial rocks by definition) to those observed on small bodies. This approach has been validated in the case of the S-type / ordinary chondrites connection by the Hayabusa mission (Nakamura et al., 2011; Nakashima et al., 2013; Tsuchiyama et al., 2014; Yurimoto et al., 2011). However, this approach requires that pieces from the surface of small bodies can be found within the suite of available extra-terrestrial materials. In the case of the most primitive small bodies, the surface material is expected to be brittle and fine-grained (< 1 μm), and its survival while crossing the Earth's atmosphere is a challenge. The meteorite approach can be applied to asteroid surfaces covered by rocks but is not applicable to surfaces covered by fragile fine-grained material.

There are numerous indications that the surface of comets and primitive asteroids (P- and D-type (Vernazza and Beck, 2016) is covered by porous dust particles made of individual grains with sizes smaller than visible wavelength (Emery et al., 2006). If the dust from which these bodies has been inherited from the interstellar medium with no further evolution in the proto-solar disk, it should consist in sub-micrometric-sized grains according to the observed interstellar extinction (Gail and Hoppe, 2010). Inter-planetary dust particles (IDPs) collected on Earth, which may come from comets or C-, P- and D-types asteroids Vernazza et al., 2015, are made by agglomeration of individual sub-micrometric grains (Rietmeijer, 1993) with macro-porosity ranging from 4 to 96 %.

Moreover, cometary dust particles collected *in situ* by the Stardust (from comet 81P/Wild 2) and Rosetta (from comet 67P) space missions are made of hierarchical agglomerates of subunits having sizes down to hundreds of nanometres and probably even smaller, as measured by the Micro-Imaging Dust Analysis System (MIDAS) instrument (Riedler et al., 2007) on board Rosetta (Bentley et al., 2016; Levasseur-Regourd et al., 2018; Mannel et al., 2019, 2016; Price et al., 2010). This hierarchical structure leads to different morphologies of the agglomerates ranging from compact (porosity < 10 %) to porous (10-95 %) or fluffy (> 95 %) particles (Güttler et al., 2019). Highly porous cometary dust particles may contribute to the porosity of the surface and of the interior of the nucleus estimated from 60 to 80 % by the Comet Nucleus Sounding Experiment by Radiowave Transmission (CONSERT) radar on comet 67P (Herique et al., 2016; Kofman et al., 2015). Besides comets, the mid-infrared emission spectra, thermal inertia and radar albedo of C and D-type asteroids are all consistent with the surface of these objects being highly porous (Emery et al., 2006; Vernazza et al., 2012) .

Reflectance spectra of the surface of comet 67P measured from 0.2 to 5.1 µm by the Visual InfraRed Thermal Imaging Spectrometer (VIRTIS) (Coradini et al., 2007) on board Rosetta revealed a geometric albedo of about 6 % at 0.55 µm (Ciarniello et al., 2015), a moderate red slope and an absorption band with low contrast around 3.2 µm (Capaccioni et al., 2015). Although the absorption band has been attributed to the presence of ammonium salt at the surface of the comet (Poch et al., 2020), the spectral slope and albedo remain largely unexplained (Rousseau et al., 2018).

In order to understand the spectral characteristics of these objects and correctly constrain their composition (identify their constituents and quantify their relative abundance) it is necessary to be able to separate the effects due to grain size and porosity from those due to composition. This is a general concern of surface remote sensing, but it is particularly important for the Vis-NIR reflectance spectroscopy of primitive small bodies, given their very peculiar surface texture. To this aim, it is thus crucial to understand how light propagates within a porous surface medium constituted of sub-micrometric grains: How do sub-micrometric grain size and porosity influence the spectral reflectance of a given material?

The influence of grain size on reflectance spectra has been extensively studied (Adams and Filice, 1967; Cooper and Mustard, 1999; Hunt and Vincent, 1968; Le Bras and Erard, 2003; Lyon, 1963; Mustard and Hays, 1997; Pieters, 1983; Salisbury and Eastes, 1985; Salisbury and Wald, 1992; Serventi and Carli, 2017). Several of these studies highlight the fact that the finest grains have dominant effects on the reflectance spectra of mixtures composed of various grain sizes (Pieters, 1983; Pieters et al., 1993), but only very few studies have addressed the influence of grains smaller than the wavelength on Vis-NIR reflectance spectra (Cooper and Mustard, 1999; Mustard and Hays, 1997). Mustard and Hays (1997) measured the reflectance spectra from 0.3 to 25 µm of several grain size separates of olivine and quartz, having mean grain sizes from about $21 \pm 5$ µm down to $3 \pm 1$ µm. Cooper and Mustard (1999) performed a similar work using samples of montmorillonite clay and a palagonitic soil. As grain size decreased, they observed a strong decrease of the absorption band depth (also called band strength, or spectral contrast) for all samples, attributed to a relative increase of scattering versus absorption. A drop of the level of reflectance for the finest grains ($< \lambda$) was also noted. However, both studies focused on the influence of micrometric dust at the surface of Mars and did not address the influence of sub-micrometric-sized grains found on small bodies. In an effort to understand the low reflectance and the spectral slope of comet 67P surface, Rousseau et al. (2018) produced sub-micrometric grains of silicate, coal and iron sulphide as cometary analogue materials. The reflectance spectra of these constituents revealed that when the grain size decreases down to the sub-micrometre range, silicate and coal remained bright whereas iron sulphide darkened. Theoretically, isolated absorbing grains smaller than $\lambda/\pi$ should absorb much more than they scatter to the point we can approximate them as absorbers only (Hapke, 2012), so the overall reflectance should decrease. Rousseau et al. (2018) explained this discrepancy observed for silicate and coal by more agglomeration of these grains compared to those made of iron sulphide. In mixtures, they found that the darkest constituent (iron sulphide) dominates the spectra and the way the constituents are mixed together can affect the spectra as much as particle size.

The influence of a high surface porosity on reflectance spectra has been even less studied. Conel, 1969, Salisbury and Eastes, 1985 and Salisbury and Wald, 1992 reported a loss of spectral contrast of the fundamental molecular vibration bands (also called reststrahlen bands) when fine grains are sifted to form a porous surface, compared to when the same grains are packed, forming a denser surface. To explain these observations, Salisbury and Wald, 1992 proposed that when hyperfine grains are packed next to each other, they scatter in phase, producing a coherent interference (coherent scattering), like a larger grain. Conversely, when they are separated by distances of the order of the wavelength (when porosity increases), they scatter out of phase as individual grains, incoherently (incoherent scattering), leading to a decrease of the spectral contrast (Salisbury and Wald, 1992; Feynman, 1964). This could also explain the observation by Poch et al., (2016) of the quasi-absence of absorption bands in the Vis-NIR reflectance spectrum of a surface made of smectite phyllosilicate having a porosity larger than 90% that was obtained from the sublimation of a mixture of water ice and smectite. Sublimation of icy-dust mixtures is a process occurring on the surfaces of comets (Pommerol et al., 2015) and on some asteroids (Hsieh, 2020), which can generate high surface porosity. However, to date there is no systematic measurement of the influence of porosity on the Vis-NIR reflectance spectra of hyperfine grains made of various materials representative of small bodies.

In the present study, we aim at improving the knowledge of how the Vis-NIR radiative transfer takes place inside hyperfine and hyperporous particulate surfaces, with the final goal to test and improve numerical radiative transfer models and provide better constraints on the composition of comets and related asteroids. To this aim, building upon the previous works from Rousseau et al. (2018) and Poch et al. (2016), we have developed experimental protocols to produce powders made of sub-micrometre-sized grains and to generate highly porous surfaces made of these hyperfine grains, reproducing the fluffy microstructure of aggregates observed in cometary dust. We then measured the Vis-NIR reflectance spectra (from 0.6 to 4.0 μm) of these samples, revealing the relative influences of grain size and porosity.

The dust covering the surfaces of comets and related asteroids appears to be mostly made of anhydrous silicates, organic matter, and opaque minerals, the latter being the most probable responsible for their low albedo from the visible to the infrared wavelengths (Quirico et al., 2016, Rousseau et al., 2018). As Rousseau et al. (2018) showed, the spectral reflectance of mixtures of these constituents is controlled by a multitude of parameters (mixing modalities, etc.) whose effects are difficult to disentangle. So, in order to improve our understanding of the radiative transfer in hyperfine and hyperporous particulate surfaces, it appears necessary to perform a systematic study of the scattering properties of each constituent taken separately, before studying mixtures.

Therefore, this first study focuses only on relatively bright silicates like olivine, pyroxene, quartz and smectite. On comet 67P, anhydrous mafic silicates are the main constituents of the mineral phase, which could represent about 55 wt% of the dust (Bardyn et al., 2017). In future studies, we will address the cases of opaque constituents and mixtures, following the same methods presented here.

This manuscript is organised as follow. In section 2, the experimental protocols enabling the production of hyperfine and hyperporous surfaces are detailed (grinding, sieving, size distribution, sublimation of ice-dust mixtures) and the methods to measure the reflectance spectra are presented. Section 3 presents the reflectance spectra measured for large-grained materials (3.1), hyperfine and compact samples (3.2) and finally hyperfine and hyperporous samples (3.3). In section 4, we discuss the effects of grain size (4.1) and porosity (4.2) on the spectra, and their implications for the interpretation of cometary and asteroid observations (4.3). Finally, section 5 summarizes the main results and findings.

## 2. Methods

### 2.1. Sample Preparation

#### 2.1.1. Samples

Stardust collected particles, Interplanetary Dust Particles (IDPs), micrometeorites that are supposed to be of cometary origin, as well as ground and satellite observations are all consistent with the presence of Mg-rich silicates and pyroxenes in cometary dust (Crovisier et al., 1996; Dobrică et al., 2012; Engrand et al., 2016; Zolensky et al., 2006).

Therefore, we chose to use a Mg-rich olivine and a pyroxene for the preparation of our samples. The Mg-rich olivine is a forsterite ($Fo = \frac{[Mg]}{[Mg] + [Fe]} \geq 94$) purchased from Donghai Pellocci Crystal Products, China. X-ray diffraction (XRD) measurements indicate trace amount of quartz (**Supplementary Figure 8**). The pyroxene sample comes from a mine in Brazil (Britt et al., 2019). XRD measurements indicate it contains enstatite (~44%) and diopside (~11%), but also a non-negligible fraction of talc (~27%) and plagioclase (9%) (**Supplementary Figure 8**). We also studied a natural rock sample collected, containing mainly smectite (or illite-smectite), but also feldspar and quartz as measured by XRD (**Supplementary Figure 8**). We chose to study this smectite-rich rock because its reflectance spectrum presents deep absorption bands due to smectite in the studied spectral range, so this sample can serve as a test to observe the spectral changes with decreasing size of grains and increasing porosity. Finally, we also used a sample of synthetic quartz spheres. We chose to include quartz in this study because unlike olivine, pyroxene and the smectite-rich it presents relatively few absorption bands in the Vis-NIR, enabling to study more precisely the effect of grain size on the spectral slope and the level of reflectance of the continuum. XRD measurements revealed that this quartz is almost entirely amorphous, we will thus refer to this sample as silica (**Supplementary Figure 8**).

### 2.1.2. Preparation of hyperfine grains

*Grindings*

To explore light propagation in cometary surfaces composed of sub-micrometre sized grains, we established a specific grinding protocol to obtain sub-micrometre sized powders from raw material of several millimetres. This new protocol results from the optimization of the one presented in Rousseau et al. (2018).

Each mineral sample is successively dry- and wet-ground several times using a Planetary Grinder Retsch© PM100, as schematized in **Figure 1**. Grinding is achieved by using zirconium oxide ($ZrO_2$) balls, of progressively decreasing sizes. It has been observed that the balls have to be at least three time larger than the grains in order to transfer enough energy to crush the materials. Usually for the first grinding, 15 mL of the raw material is put into the bowl with the 15 mm diameter balls of $ZrO_2$ for 20 minutes of dry-grinding. The obtained powder is then passed through sieves of 400, 200, 100, and 50 µm. The fraction of the powder smaller than 50 µm is stored for the last step of the grinding protocol. The fractions of the powder larger than 50 µm is then dry-grinded again with smaller balls (2 mm) until we obtain at least 15 mL of grains finer than 50 µm, which are necessary for the next steps. The powder smaller than 50 µm is then wet-grinded using ultra-pure ethanol and balls of 500 µm for 150 minutes. Ethanol is used to maintain the grains in suspension and improves the efficiency of grinding. To establish and validate the protocol, the mixture in the grinding reactor was recovered, and the size of the grains measured, at different grinding times to determine the optimum time of grinding in order to obtain hyperfine grains. This optimum time depends on the hardness of the mineral sample. Soft samples such as the smectite-rich mineral mixture which, in our case, contains a lot of smectite, shows a significant amount of hyperfine grains after only 20 minutes of colloidal grinding. The silica that we use presents a medium hardness, and it becomes hyperfine after around 40 minutes of colloidal grinding. Olivine on the contrary is harder to grind down to sub-micrometre size, and 150 minutes of colloidal grinding were required to obtain a majority of grains smaller than 1 μm, as shown in **Table 1**.

After the grinding, the mixture "balls + ethanol + powder" is sieved through mesh of 25 μm to separate the powder from the balls and to eliminate big grains (> 25 μm) that could have escaped fragmentation. Note that grains of a few micrometres are present in the final powder, but they represent a minority of the total number of grains (less than 1%; see **Figure 4** and figures in Supplementary Materials). Also note that sieving the powder with ethanol optimize the recovery of the sample compared to a dry sieving. The final powder recovery was first achieved by evaporating ethanol under a fume hood or in a stove. This procedure is however time consuming and inappropriate for samples that are easily oxidized. We then turned to another way which consists in a centrifugation of the colloidal solution at 2000 rpm. The mud-like residue was then dried in a cold evaporator, under primary vacuum. The dried powder was then recovered by scratching the tubes and stored in a desiccator.

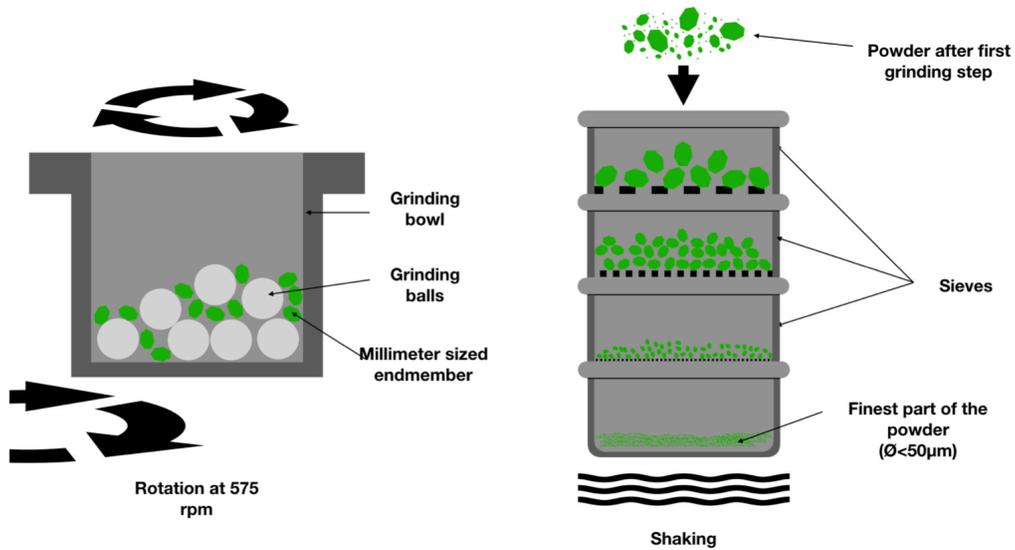

**(a) Dry grinding and sieving**

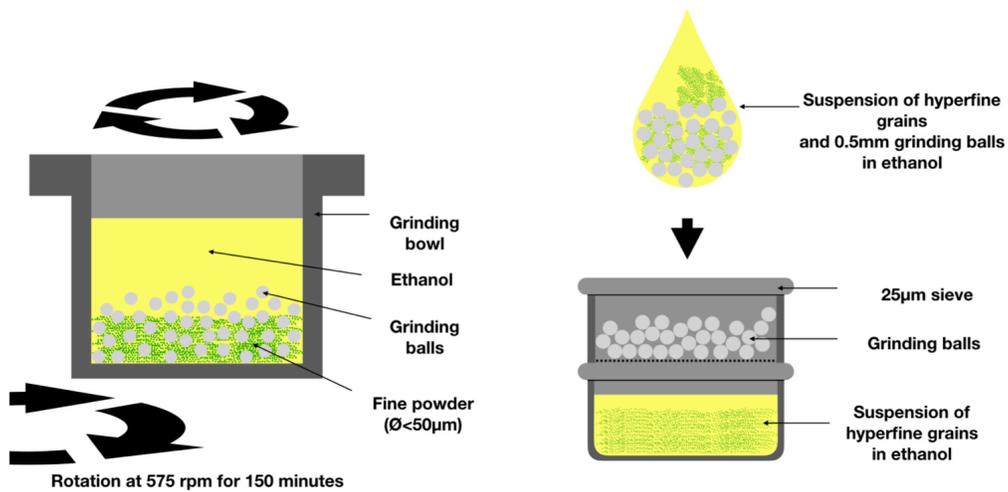

**(b) Wet grinding and sieving**

**Figure 1:** Diagram of the grinding and sieving steps necessary to prepare hyperfine (sub-micrometre sized) grains from millimetre-sized materials: **(a)** the first dry grindings and sieving steps to get fine powder smaller than 50 μm starting from millimetre-sized material, and **(b)** the wet grinding and sieving steps to get hyperfine (< 1 μm) powder from the powder smaller than 50 μm.

*Measurements of the grain size distribution*

Grain size distribution has been determined by performing Scanning Electron Microscopy (SEM) imaging. SEM images were obtained in Grenoble at the Consortium des Moyens Technologiques Communs (CMTC) using a SEM Zeiss Ultra 55. **Figure 2a** shows SEM images of the olivine grains.

Particle size determination is made by hand-counting and contouring particles using ImageJ software (**Figure 2b**). To have a good statistic of grain size distribution, we analysed at least 1000 grains on each SEM image. Size is estimated by fitting an ellipse to each particle contoured shape and the semi-major axis of the fitted ellipse is kept as the particle size. The fit is made by adjusting the area, the centre and the orientation of the ellipse on the selection (contoured shape of the grain), as seen. When the grains have a complex shape, as seen on the right of the image of **Figure 2b**, the fit can underestimate the real size of the grain. This method presents the advantage to quantify the projected 2D area of the grain, which can be used as a cross section for modelling purpose.

To count and measure each grain on SEM images, grains must be well dispersed and isolated and not forming aggregates on the sample holder. Dispersing grains on the surface of the SEM sample holder also prevents from biases due to overlapping grains that can hide the smallest ones and thus increase the mean value of particle diameter in the grain size distribution. To this aim, a tiny spatula of the powder has been put in 5 mL ethanol and agitated by ultrasonication to destroy the aggregates using a Hielscher 200Ht ultrasonic unit equipped with a 7 mm diameter sonotrode. A small drop of the mixture was then deposited on the SEM sample holder. Grain size distributions determined for several grinding times and samples are available in Appendix; all mean values and largest size observed for these samples are presented in the **Table 2**. With increasing grinding time, all these samples are showing a diminution of their mean grain size, and more importantly a reduction of their maximum grain size.

As shown on **Figure 2c**, the mean value of the particle size distribution is about 0.69 ± 0.47 µm for olivine. This size is well within the range of the measurements made by MIDAS/Rosetta instrument (Bentley et al., 2016; Mannel et al., 2019, 2016) where grains sizes lie between 0.1 and 4 µm.

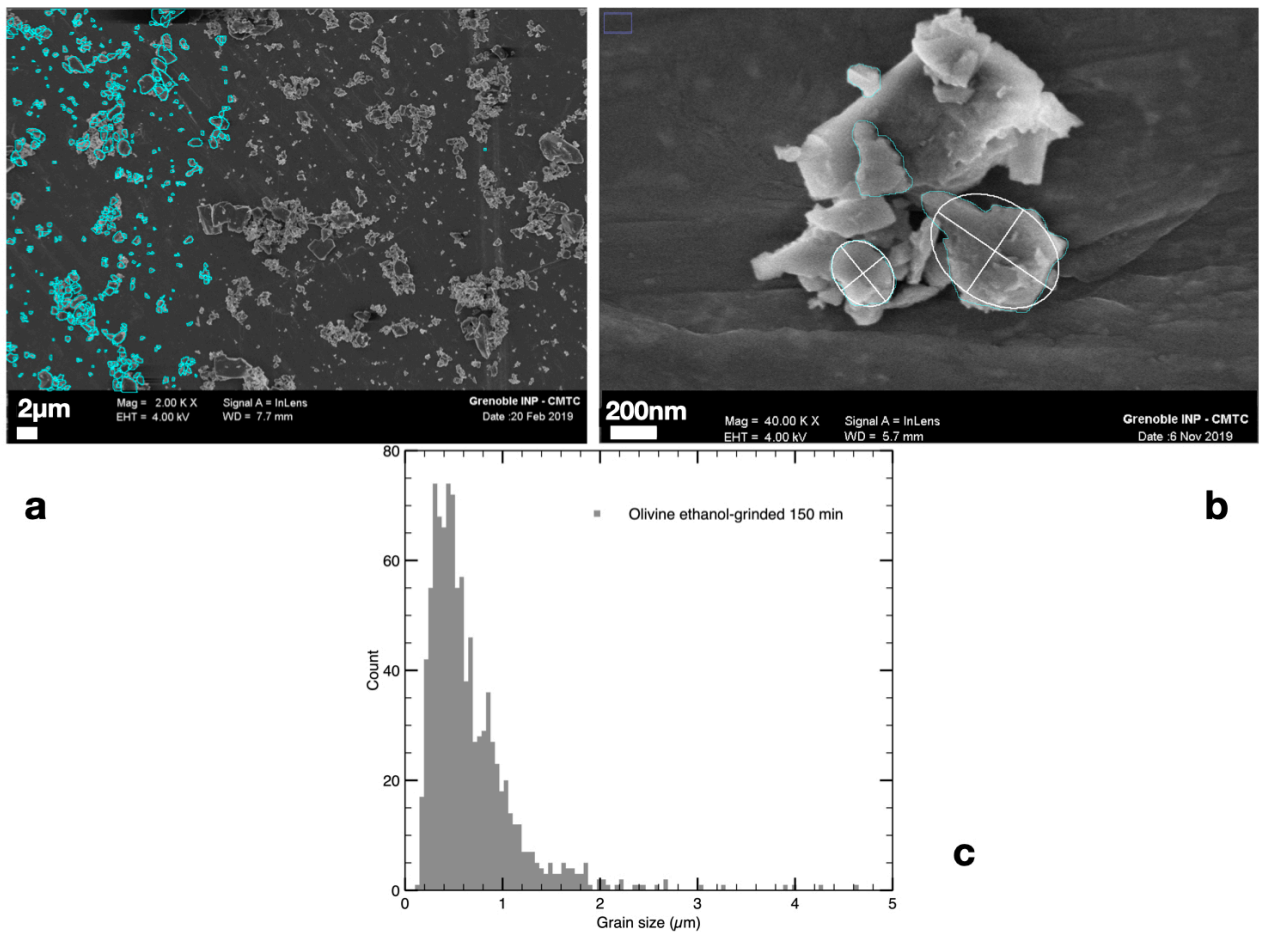

**Figure 2:** Determination of the grain size distribution: **(a)** Scanning Electron Microscopy (SEM) image of olivine grains after 150 min wet-grinding. Part of these grains were counted, contoured by hand, and ellipses were fitted to their contours to estimate their size. The blue lines represent the contours of the grains that have already been contoured and counted; this image was intentionally made during the counting to show the method and the raw SEM image. **(b)** SEM image of a small aggregate. This is an example of the method used to obtain estimation of the grain size. Ellipse area are adjusted using a least-squared method on the contoured shape of grains. **(c)** Size distribution of olivine grains, hand-counted from the SEM image, as a function of the major axis of the fitted ellipse. The mean size of this distribution is ~0.4 µm, showing the validity of the grinding protocol. The size distributions of the other mineral samples and those measured to validate this grinding protocol are available in Supplementary Materials.

| | **Mean grain size (µm) \| Largest size (µm)** | | | |
|---|---|---|---|---|
| **Wet Grinding time (min)** | 20 | 40 | 60 | 150 |
| Olivine | 1.47±1.61 \|16.59 | 1.53±1.51 \|11.59 | 0.69±0.71 \|4.48 | 0.69±0.47 \|4.67 |
| Pyroxene | - | - | - | 0.41 ± 0.168 \|2.28 |
| Silica | 1.5 ± 0.63 \|4.87 | 0.54 ± 0.46 \|2.08 | 0.32 ± 0.35 \|2.39 | - |
| Smectite-rich | - | - | - | 0.18 \| 1.15 * |

**Table 1 :** Grain size after colloidal grinding as a function of grinding time. All the size distributions are available in the Supplementary Materials. * Because of their disk shape, the presented value corresponds to a disk diameter, but it does not take in account of the thickness of the grains.

### *2.1.3. Preparation of hyperporous surfaces: sublimation residues*

Several laboratory experiments performed since the 70s have shown that when water ice mixed with impurities sublimates under cometary-like conditions ($10^{-5}$-$10^{-6}$ mbar and 180-240 K) solid ice-free structures are formed after sublimation. These structures, called "sublimation residues", can exhibit high porosities up to about 90% (Dobrovolsky and Kajmakow; 1977, Poch et al., 2016; Saunders et al., 1986, Pommerol et al., 2019). In the present study, we produced hyperporous granular surfaces following the same experimental protocol described in Poch et al. (2016) and summarized below.

We only used the smallest grain separates, i.e. the hyperfine grains, to produce the porous surfaces mainly because these grains are more representative of those present on primitive small bodies. The production of the hyperporous surfaces was done in two steps.

The first step was the production of ice particles containing the grains, using the Setup for Production of Icy Planetary Analogues - B (SPIPA-B) detailed in Pommerol et al. (2019). The powder made of sub-micrometre-sized grains was poured into a beaker of ultra-pure liquid water and this suspension was agitated using ultrasound, to prevent formation of aggregates and sedimentation of grains.

Then, the suspension was pumped using a peristaltic pump and nebulised into a bowl filled with liquid nitrogen. As the droplets of water containing grains encounter liquid nitrogen, they froze immediately, so that grains were trapped inside the ice, forming veins between ice crystals (Poch et al., 2016).

Once the beaker was empty, the bowl containing ice and liquid nitrogen was left to evaporate in the fridge at 250 K until all the liquid nitrogen was gone. When this dirty ice was free of liquid nitrogen, it was sieved using a 400-μm sieve over a sample holder to gently depose the ice without pressing or compacting it. This cylindrical sample holder in aluminium was 5 cm in diameter and 2 cm deep.

The second step was the sublimation of the ice and the consecutive formation of the sublimation residue. The sample holder was transferred in the Carbon-IR vacuum chamber (Grisolle et al., 2014). Inside this vacuum chamber, the cell and the sample were maintained at a temperature of 173 K and under a pressure of $10^{-6}$ mbar, so that the water ice could slowly sublimate. This chamber, closed by a sapphire window and coupled to a spectro-gonio-radiometer, enabled *in situ* measurements of reflectance spectra of the sample during the whole process which lasts about 100 hours.

After complete sublimation of the water ice, only a porous and fragile residue, the so-called sublimation residue, remained inside the sample holder. At this stage, the sample holder was removed from the cold vacuum chamber and allowed to warm slowly to ambient temperature in a dry container filled with dinitrogen. Fragments of the sublimation residue were carefully collected and stored in a desiccator for further measurements.

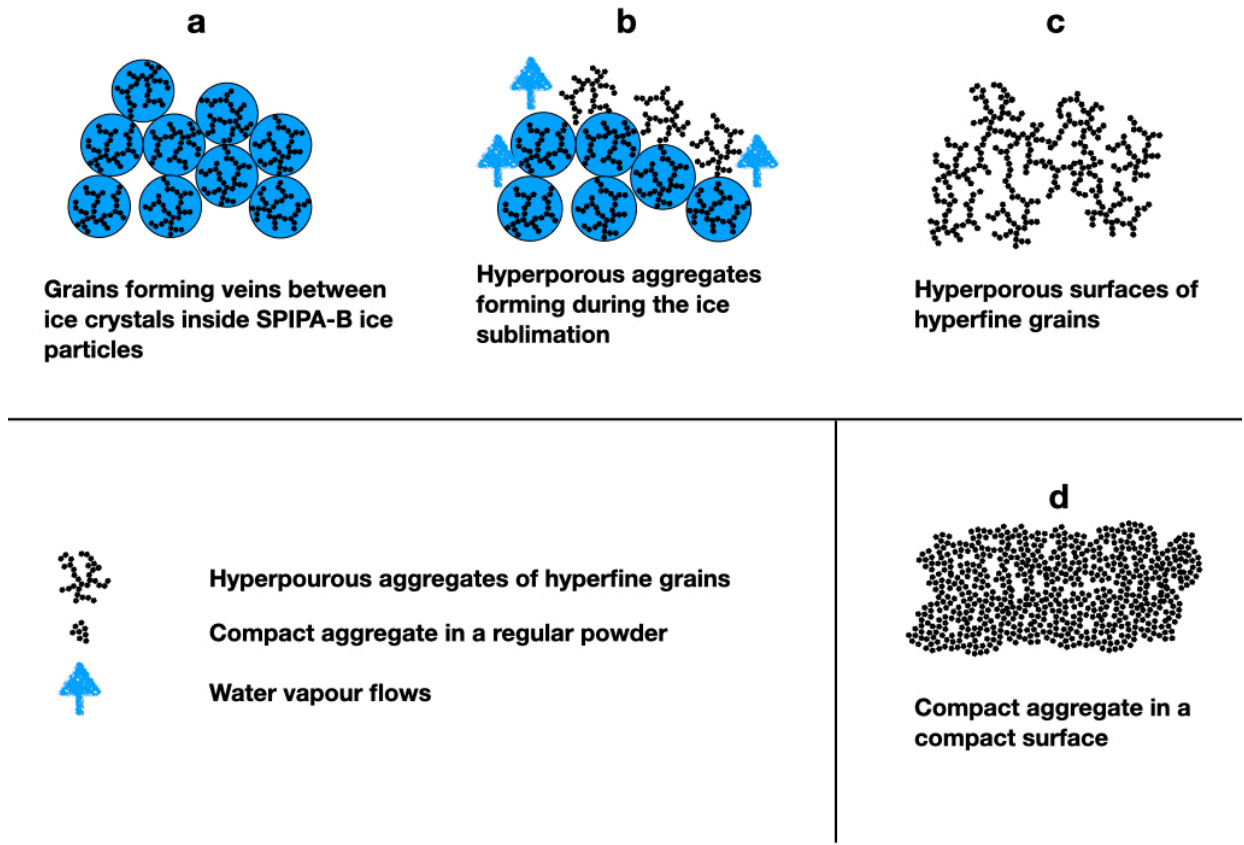

**Figure 3 :** Diagram of the sublimation process forming hyperporous surfaces. **(a)** Inside each ice particle produced using SPIPA-B, the mineral grains are forming veins between water crystals. **(b)** During the sublimation of the ice, the water crystals are replaced by voids between the mineral veins, and the water vapour may move some grains. **(c)** After complete sublimation of the ice, the grains form a porous structure (with a typical porosity of 99%), called the sublimation residue. **(d)** Compact surface with the same grain size for comparison.

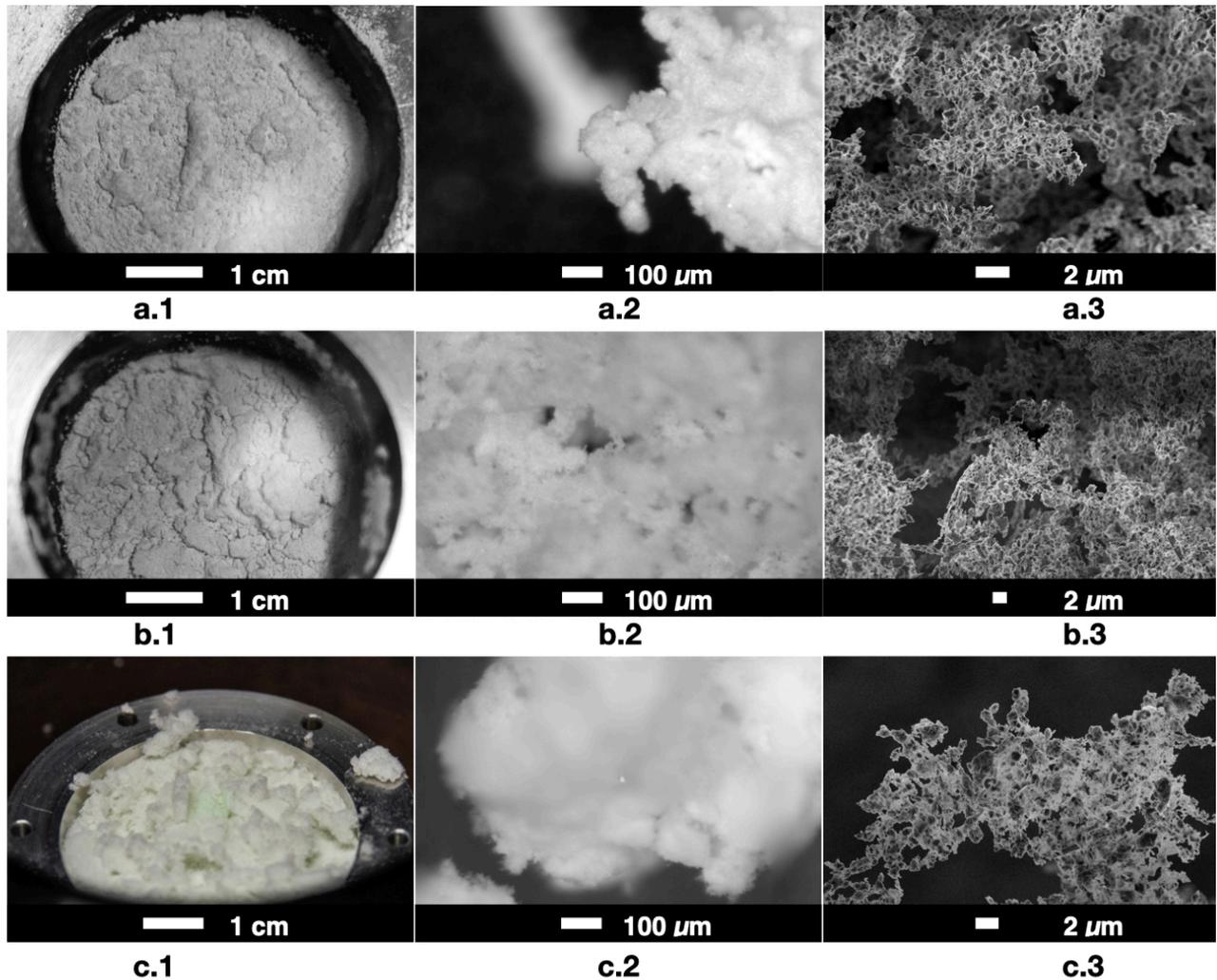

**Figure 4 :** Sublimation residues produced from **(a)** olivine, **(b)** pyroxene and **(c)** smectite-rich materials. The images have been taken at different scales using, from left to right, a simple camera, a binocular microscope, and a SEM. These images show the different textural aspects with respect to the scale. Macroscopic aspect can be seen in the images taken with a camera (first column), it shows millimetre to centimetre-sized aggregates. Binocular images (second column) highlight the largest part of the **macro-porosity** (tens to hundreds of micrometres) between aggregates. Aggregates have a woolly texture at this scale, and spikes can be discerned. SEM images (third

column) show in more details the **macro-porosity** between aggregates, and individual grains constituting the aggregates can be spotted, revealing the **micro-porosity** (tens to hundreds of nanometres) between grains. Grains are forming sheet-like aggregates and spikes of about 10 µm large where each grain is separated by less than 1 µm. Contact between two grains can be reduced to one edge only.

### Hyperporous surface

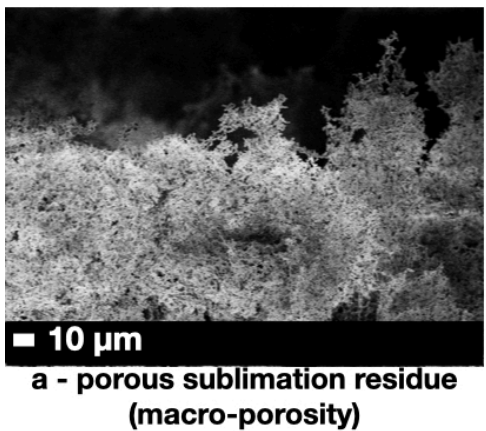

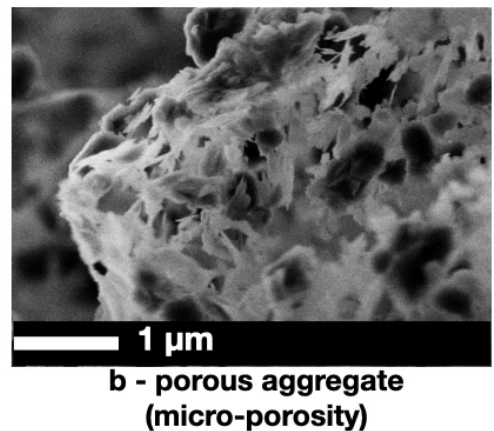

a - porous sublimation residue
(macro-porosity)

b - porous aggregate
(micro-porosity)

### Compact surface

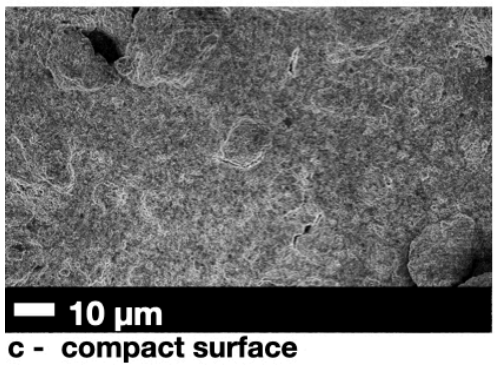

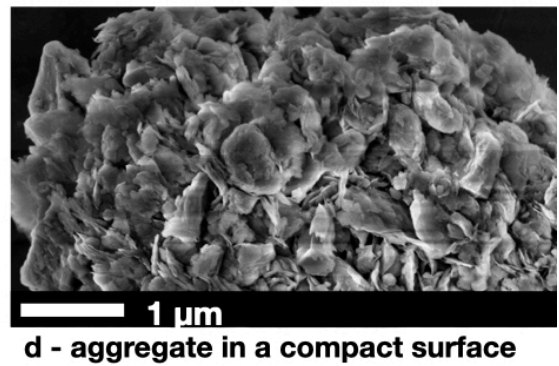

c - compact surface

d - aggregate in a compact surface

**Figure 5 :** SEM images showing hyperfine surfaces made of the smectite-rich material. **(a)** In the hyperporous surfaces (sublimation residue), the estimated porosity is around 99.5%. Grains appear to be organised in aggregates spaced by tens of micrometres, which we refer to as the "**macro-porosity**". **(b)** A close-up view on an aggregate show that the material seems to be organized in clusters with average size of hundreds of nanometres (in dark grey) surrounded a by lower density structure less than 50 nm thick (in light grey). The low-density structure is probably made of smectite platelets, and the denser clusters of the other minerals (feldspar, amorphous silica). This image reveals spaces of tens to hundreds of nanometres between individual grains/platelets inside a fluffy aggregate, which we refer to as the "**micro-porosity**". **(c)** Image of a compact surface of hyperfine grains of smectite. **(d)** Composite image of a compact aggregate that can be found in a compact surface of the smectite-rich material.

## 2.2. Reflectance measurements

We measured the reflectance spectra of the samples using the SpectropHotometer with variable INcidence and Emergence (SHINE) at IPAG (Brissaud et al., 2004) working in micro-beam mode (as described in section D.2 of Potin et al., 2018). All the samples have been studied under the same geometry: sample surfaces were illuminated on an area of 7.5 mm in diameter by a tuneable monochromatic light source at normal incidence, and the observation angle was set to 30°, resulting in a phase angle of 30°. Two detectors measured the light reflected by the samples in the observation direction over the wavelength range from 0.4 to 4.2 μm. To obtain absolute values of reflectance spectra, the detected signal was calibrated using surfaces made of Spectralon® (in the interval 0.4-2.0 μm) and Infragold® (in the interval 2.0-4.2 μm) as references (Labsphere Inc.). Using these reference measurements, and after applying corrections for the small non-lambertian behaviour of the Spectralon® (Bonnefoy, 2001), we computed the Bidirectional Reflectance Factor (BRF), which is the ratio of the flux reflected by a surface to the flux reflected by an ideal lambertian surface. Spectra measured under vacuum inside Carbon-IR were corrected from the contribution of the sapphire window.

This window induces multiple reflexions between the sample and the window itself. The correction, developed in Pommerol et al., 2009, is based on the fact that the flux detected by the sensor is the sum of the flux reflected by the window, the flux reflected by the sample that goes back through the window, as well as multiple scattering between sample and window that make their way out through the window. This model uses optical index and transmission properties of sapphire.

Compact samples were prepared by depositing the powder inside the sample holder and levelling their surface with a straight spatula.

## 2.3. Spectral metrics

### 2.3.1. Band depths

All along this study, we will use the term band depth to indicate the relative band depth (expressed as %), computed as follow:

$$Band\ depth\ (\%) = (1 - \frac{R_b}{R_c}) \times 100$$

where $R_b$ is the minimum reflectance inside the band and $R_c$ the reflectance of the continuum interpolated at the same wavelength. $R_c$ was estimated by drawing a line between two points at the edges of the band and taking the value on the line at the wavelength of the minimum of the band, as shown in **Figure 6**.

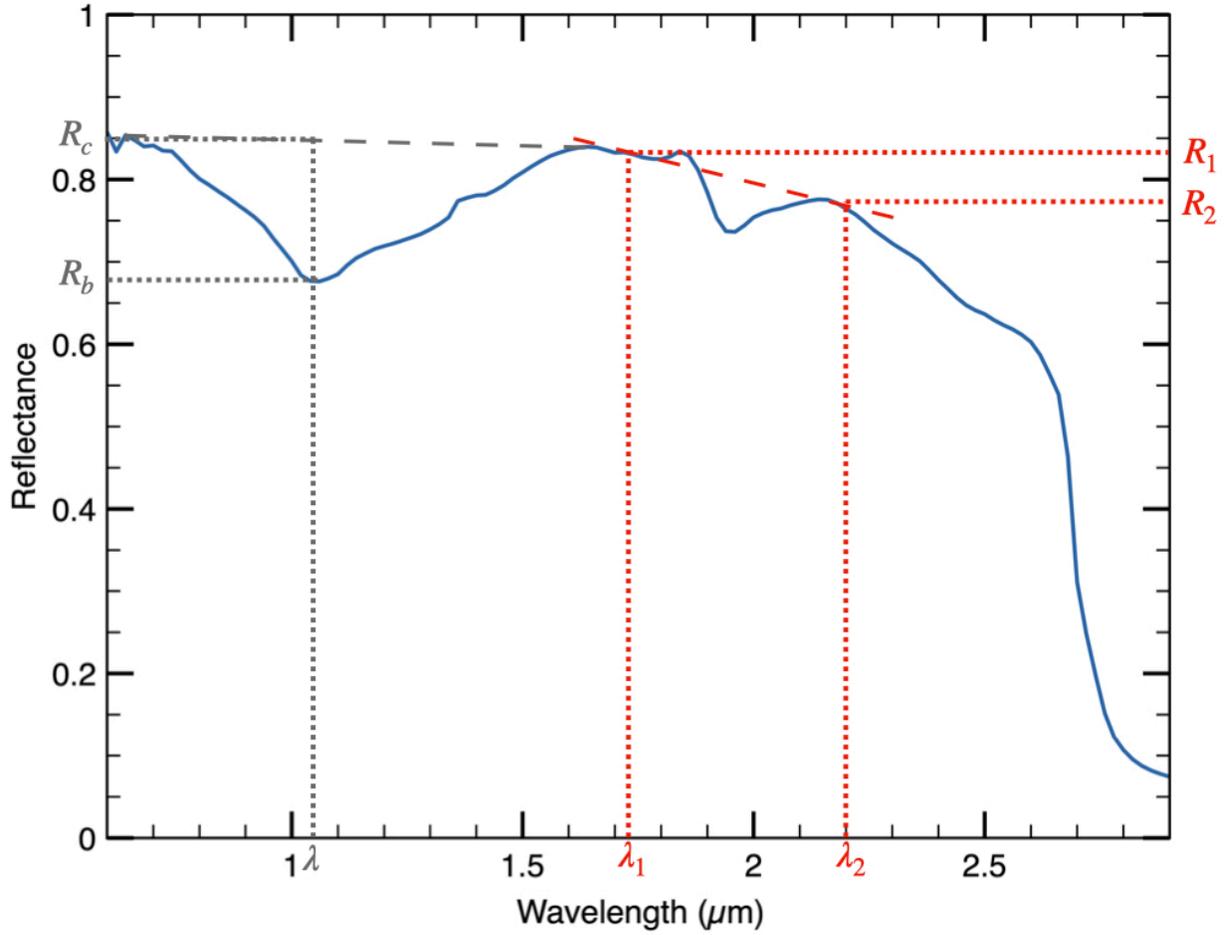

**Figure 6 :** Example of computation of band depth on the 1-μm band of olivine and of spectral slope. Two points on the edges of the band define the continuum. $R_c$ is the reflectance of the continuum at the spectral position of the band minimum $R_b$. Two points $R_1$ and $R_2$ on the continuum and their wavelength are used to compute the spectral slope.

### 2.3.2. Spectral Slope

Spectral slope is computed following Fornasier et al. (2015) and Delsanti et al. (2001)

$$Spectral\ slope\ (\%.\,(100\ nm)^{-1}) = \frac{R_2 - R_1}{R_1 \times (\lambda_2 - \lambda_1)} \times 10^4$$

where $R_n$ is the reflectance of the continuum at the wavelength $\lambda_n$ (in nanometres). Because of the presence of deep bands below 1.5 μm and from 2.8 μm in each spectrum, spectral slopes are difficult to estimate. We chose to adapt the slope computation interval outside of the absorption features. Hence the slopes are computed in the near infrared part of the spectra around the water band at 1.9 μm. However, around this water band at 1.9 μm, some absorption due to O-H bond related to the presence of water vapour on the optical path occurs. This absorption results in light increases or decreases of the reflectance between 1.8 and 1.9 μm, but also in the 1.3-1.4 μm and 2.5-2.9 μm regions. We adapted the slope computation range with respect to these water vapour absorptions.

# 3. Results

## 3.1. Reflectance spectra of large-grained initial materials (100-200 μm)

**Figure 7** presents the reflectance spectra of the prepared surfaces, constituted of grains of each material (olivine, amorphous silica, smectite-rich material, pyroxene) with decreasing grain sizes from 100-200 μm to sub-micrometre.

In all these spectra, a strong absorption feature lies around 3 μm. It corresponds to the absorption of light by water adsorbed on the grains as well as water constitutive of the materials. As the grain size decreases, the specific surface area of the sample increases, and more water vapour can be adsorbed, leading to stronger absorption features above 2.8 μm (Potin et al., 2020). Therefore, around these wavelengths it is difficult to disentangle if the changes of the spectra are due to a new regime of radiative transfer or to an increasing amount of adsorbed water. Controlling the amount of this adsorbed water is difficult and it was not attempted in the present study. Consequently, spectra on **Figure 7** are only presented up to the 2.5 µm.

In this section, we focus on the reflectance spectra of the surfaces made of the largest grain sizes from 100 to 200 µm (dark blue spectra in **Figure 7**).

*Olivine*

The blue spectrum (100-200 µm) shown in **Figure 7a** presents a deep absorption feature around 1 µm due to the presence of $Fe^{2+}$ in the mineral. This structured band with 2 shoulders is in fact the combination of 3 bands related to iron. This band will be studied as an indicator of spectral changes. Some small absorption features can also be seen in the 2.2-2.5 µm region. They are more likely related to a minor contamination by phyllosilicates. Olivine powder presents a relatively high reflectance in the Vis-NIR, with a maximum approaching 70%.

*Pyroxene*

Two absorption bands are present in the dark blue spectrum (100-200 µm) shown in **Figure 7b**, around 0.9 and 1.8 µm also due to the presence of $Fe^{2+}$ in the mineral. We will focus on these two bands to follow spectral changes occurring when the grain size is decreased. The pyroxene powder is relatively dark in the Vis-NIR, with a maximum reflectance around 30 %.

*Smectite-rich material*

The spectrum shown in **Figure 7c** has a strong absorption feature composed of several bands between 0.5 and 1.5 µm, due to the presence of $Fe^{2+}$ and $Fe^{3+}$ in phyllosilicate and to $OH/H_2O$. Other bands due to $OH/H_2O$ are present at 1.9 and 2.2 µm. The powder of the smectite-rich material has an intermediate level of reflectance in the Vis-NIR, around 50 %.

*Silica*

The spectrum shown in **Figure 7d** displays spectral signature of $OH/H_2O$ at 1.4, 1.9, 2.45 and 2.7 µm. The 2.2-2.5 µm region also displays some features which can be related to the absorption of light by Si-OH bonds. Silica powder is bright in the Vis-NIR, with a visible reflectance around 90 %. The powder appears actually white and bright when looking at it with naked eye.

## 3.2. Reflectance spectra of silicates with a decreasing grain size (25-100 µm)

**Figure 7** shows the changes occurring in the reflectance spectra when the grain size of the different materials is decreasing but still several times larger than 1 µm. We observe qualitatively for each sample:

- an increase of the reflectance level on the whole spectral range,
- a decrease of the band depth. Absorptions bands are shallower for small grains (25-100 µm) than for larger (100-200 µm),
- a decrease of the spectral slope (spectral blueing) especially for the largest wavelength (in the 2-2.5 µm region).

## 3.3. Reflectance spectra of hyperfine silicate samples (< 1 µm grain)

In **Figure 7** we can notice the similar observations as for grains in the 25-100 µm size range, but more pronounced. Hence, we observe:

- an increase of the reflectance for the shortest wavelengths, although we can observe a decrease in the 1.5-2.5 µm region,

- a strong decrease of the band depth. Powders constituted by sub-micrometre-size (sub-µm) grains present much shallower bands with respect to the powder containing much larger grains. The most striking examples are the spectra of olivine and pyroxene (resp. **Figure 7a**, **Figure 7b**),

- spectra of hyperfine powders show a much bluer slope than those of powders with large grains (the most noticeable examples are pyroxene and silica **Figure 7b**, **Figure 7d**).

The spectral blueing occurring for all powders when the grain size decreases is well visible in **Figure 8b**. This plot exhibits the ratio between the spectrum of the powders with small grains (< 50 µm) and those with sub-µm grains.

For each material, this ratio emphasizes the reduction of band depth and the change of spectral slope (**Figure 8**). **Figures 4** to **7** in Supplementary Materials show these ratios for each material.

Inspection of **Figure 8** shows that the slope of these spectral ratios (i.e. the spectral blueing) is similar for each material. Between 1.6 and 2.5 µm the ratio for olivine and silica is very similar. For the pyroxene the spectral blueing is slightly lower but still very close to what we observe with olivine and silica. Smectite-rich ratio's slope (spectral blueing) is hard to determine because of the presence of many absorption bands over most of the spectrum, but it seems analogous to the others in the continuum between 1.6 and 1.8 µm and between 2.1 and 2.3 µm.

Such effects could be explained by an increase of the amorph phase during the grindings, reducing the crystalline fraction, leading to a decrease of the band contrast. We checked the presence and the shape of the silicate feature in the mid-infrared around 10 µm by measuring a pellet of KBr containing a hyperfine powder of olivine (Supplementary Figure 9). The silicate feature is present and sharp, indicating that the fraction of amorph olivine is low in the powder. Even if it does not prove that the grinding protocol does not amorphize the sample, it clearly indicates that the band contrast decrease is more likely driven by the decrease of the grain size.

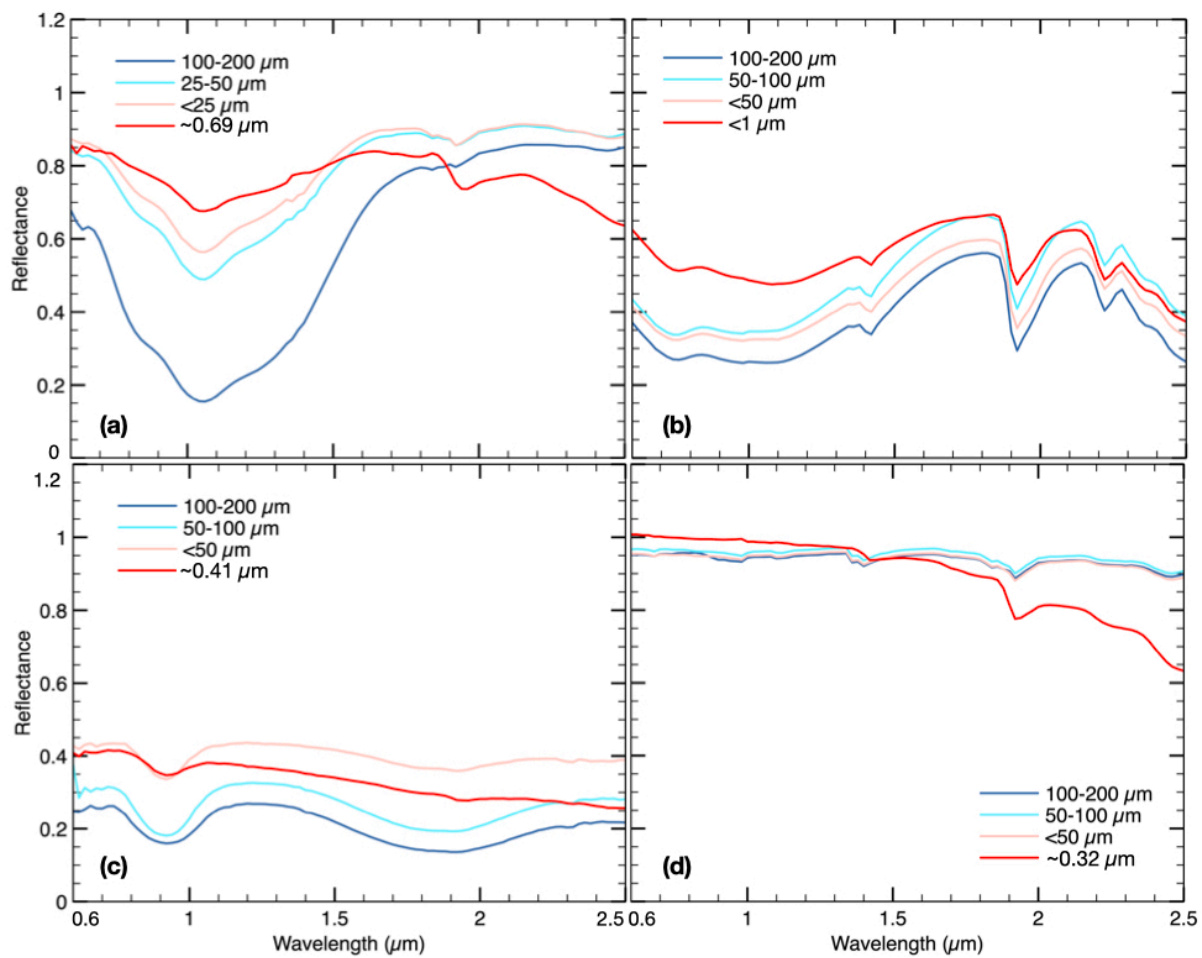

**Figure 7 :** Reflectance spectra of powders of **(a)** olivine, **(b)** pyroxene, **(c)** smectite-rich and **(d)** silica having different grain sizes from 200 µm down to sub-micrometre. This figure shows how

the grain size influences the shape, the slope and the band contrast of reflectance spectra of granular surfaces.

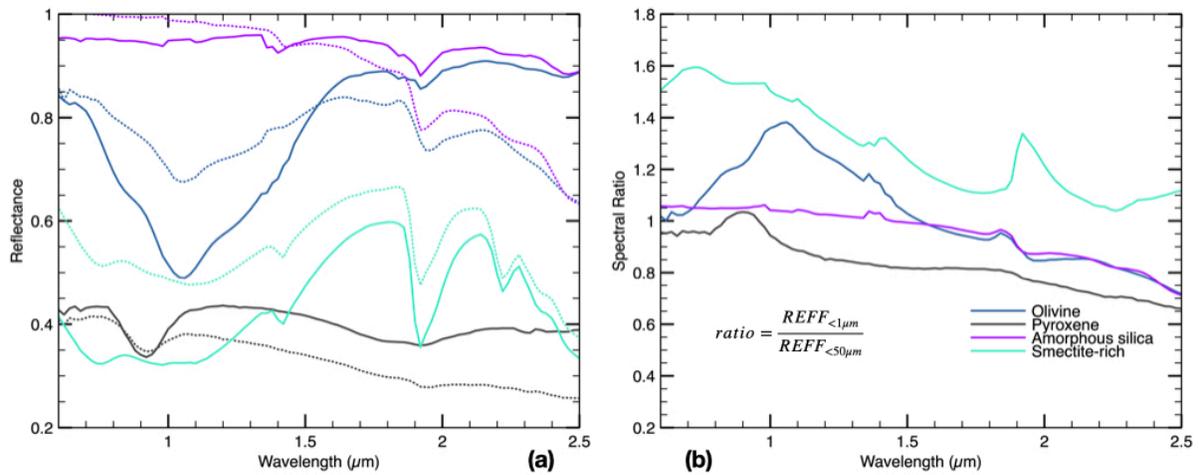

**Figure 8 : (a)** Reflectance spectra of surfaces made of grains smaller than 50 µm (solid lines) versus surfaces made of sub-micrometre-sized grains (dotted lines), for each material (olivine, pyroxene, silica and smectite-rich). **(b)** Spectral ratios computed by dividing the spectrum of the < 50 µm grains by the spectrum of the hyperfine grains, for each material. These spectral ratios highlight the changes occurring in the spectra when the grain size is decreasing down to the sub-micrometre. All these ratios exhibit a negative slope from 0.6 to 2.3 µm, indicative of a spectral blueing occurring for all samples when the grain size decreases down to the sub-micrometre. This bluing appears to be qualitatively similar for all the materials, although this effect may be more difficult to observe for the smectite-rich sample because of the presence of many broad absorption features. Ratios plotted individually are available in Supplementary Materials.

## 3.4. Structure and reflectance of hyperfine and hyperporous silicate samples

### 3.4.1. Morphological description of the sublimation residues

The structures inside the sublimation residues were observed by SEM imaging. All the residues display a variety of forms and fluffiness at the micrometre scale.

**Figure 4** and **Figure 5** present examples of the fluffy structures and high porosity that can be achieved in sublimation residues. The porosity is estimated by determining the volume occupied by the residue in the sample holder after the sublimation. It is difficult to estimate, but we found that the porosity of sublimation residues lies around 99 %.

The mechanical resistance of hyperporous media is likely dependent on the grain size. Indeed, very fine grains present thinner edges, so that short range interactions between grains will be stronger and will help the cohesion of the ensemble. Also, finer grains are lighter than larger grains so the ratio between electrostatic and gravitational interactions allows the preservation of a high-porosity residue. But for larger grains ($\geq$ 50 µm), gravity is probably stronger than the short-range bonds and our experience has shown that samples tend to collapse, reducing the porosity.

These sublimation residues are showing interesting structures such as fine filaments or planes (**Figure 4**a).

As seen in previous work (Poch et al., 2016; Saunders et al., 1986) and schematised in **Figure 3**, when the ice crystals are formed, the non-ice grains are segregated at interfaces between ice crystals. During the sublimation, these structures can disappear, be modified by the water vapour flows or they can also keep the organization they had inside ice.

All samples present two types of porosity: a **macro-porosity** (**Figure 5a**) and a **micro-porosity** (**Figure 5b**). In contrast to materials science, we define here macro-porosity as a space of a few to tens of micrometres between the aggregates. On the contrary, the micro-porosity is present inside aggregates between the individual grains and is characterized by a spatial scale comparable with the grain size (a few tens to hundreds of nanometres).

We produced sublimation residues starting from sub-µm powders of the three materials: olivine, pyroxene and the smectite-rich.

The final samples display different macroscopic surface textures and have different behaviour in the vacuum cell, depending on the ability of the grains to create bonds to each other. Residues containing phyllosilicates, such as those of the smectite-rich material, are more akin to become very fluffy, and they tend not to collapse (**Figure 4**). However, they tend to explode if the sublimation rate is too high (if the sample is heated too rapidly). They also form large fluffy chunks of sublimation residues.

Sublimation residues produced only from olivine and silica are very fragile and they tend to collapse to the bottom of the sample holder, so they are supposed to display a smaller porosity than the phyllosilicate ones. They also have to be manipulated with care.

### 3.4.2. Reflectance spectra of the sublimation residues

**Figure 9** displays the spectra of hyperfine powders of olivine, pyroxene and smectite-rich materials, and the sublimation residues made from these three powders.

We are unfortunately not able to present a spectrum for the sublimation residue made from silica.

Indeed, this sublimation residue was very fragile: it collapsed in the sample holder, resulting in a very irregular surface layer preventing a reliable measurement of a reflectance spectrum.

Spectra of highly porous sublimation residues have interesting characteristics. They all tend to have shallower bands than the spectra of larger grains and to the compact hyperfine powder. The reflectance of the continuum is also affected when measuring hyperfine and hyperporous sample. The spectral evolution from hyperfine compact powder to hyperfine and hyperporous powder (**Figure 9**) is similar to the evolution seen from large to small grains as absorption bands are shallower for spectra of hyperfine and hyperporous sublimation residues. All the values describing the difference between spectra of powders and sublimation residue for olivine, pyroxene and smectite-rich are available in **Table 1**. They will be discussed in the following section.

Reflectance level of the spectra of olivine and smectite are very similar for both the compact and the hyperporous surfaces (**Figure 9a**, **Figure 9b**). However, for the hyperporous surface of pyroxene a constant increase of reflectance of about 10% is observed on the whole spectral range compared to the compact surface (**Figure 9c**). This observation may be explained by the non-negligible presence of talc inside this powder: the rearranging of talc during the sublimation might make it more visible than in the compact powder. This rearranging of phyllosilicates has already been described in (Poch et al., 2016).

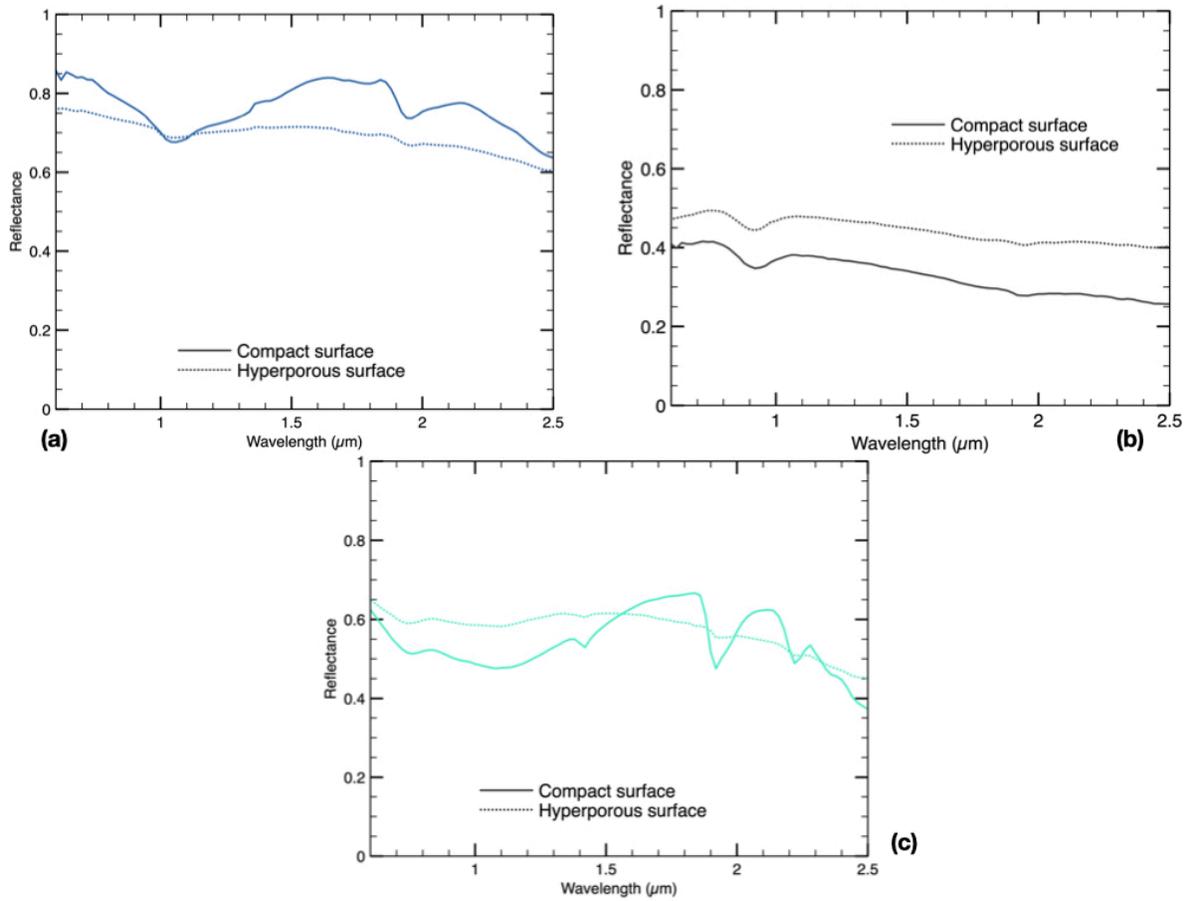

**Figure 9 :** Spectra of hyperfine grains in hyperporous sublimation residues (solid lines) or compact surfaces (dotted lines) made of **(a)** olivine, **(b)** pyroxene, and **(c)** the smectite-rich material. In each case, the hyperporous surface shows much shallower bands than the compact powder.

| | Relative band depths (%) | | Relative spectral slopes (%. 100 nm$^{-1}$) | Reflectance factor |
|---|---|---|---|---|
| **Olivine** | *at 1.0 μm* | | *from 1.74 to 2.2 μm* | *at 1.8 μm* |
| **100-200 μm** | 76.56 | | +0.238 | 0.792 |
| **25-50 μm** | 41.66 | | +0.047 | 0.884 |
| **0-25 μm** | 34.87 | | +0.025 | 0.902 |
| **sub-μm (0.69 μm)** | 19.25 | | -0.171 | 0.829 |
| **sub-μm hyperporous** | 6.72 | | -0.134 | 0.707 |
| **Pyroxene** | *at 0.9 μm* | *at 1.8 μm* | *from 1.30 to 2.36 μm* | *at 1.2 μm* |
| **100-200 μm** | 18.61 | 18.92 | -0.175 | 0.27 |
| **50-100 μm** | 19.28 | 9.71 | -0.123 | 0.33 |
| **0-50 μm** | 22.25 | 10.78 | -0.094 | 0.44 |
| **sub-μm (0.41 μm)** | 13.10 | - | -0.246 | 0.37 |
| **sub-μm hyperporous** | 8.96 | - | -0.121 | 0.75 |
| **Smectite-rich** | *at 0.9 μm* | *at 1.9 μm* | *from 1.82 to 2.14 μm* | *at 1.8 μm* |
| **100-200 μm** | 43.42 | 45.94 | -0.152 | 0.67 |
| **0-50 μm** | 36.44 | 39.00 | -0.124 | 0.67 |
| **sub-μm** | 28.52 | 25.89 | -0.203 | 0.67 |
| **sub-μm hyperporous** | 7.49 | 3.51 | -0.258 | 0.59 |
| **Quartz** | | | *from 1.72 to 2.2 μm* | *at 1.65 μm* |
| **100-200 μm** | | | -0.039 | 0.87 |
| **50-100 μm** | | | -0.046 | 0.95 |
| **0-50 μm** | | | -0.049 | 0.97 |
| **sub-μm (0.32 μm)** | | | -0.305 | 0.93 |

**Table 2 :** Table reporting the values measured for the band depths, the slopes and the reflectance

of NIR continuum of all samples

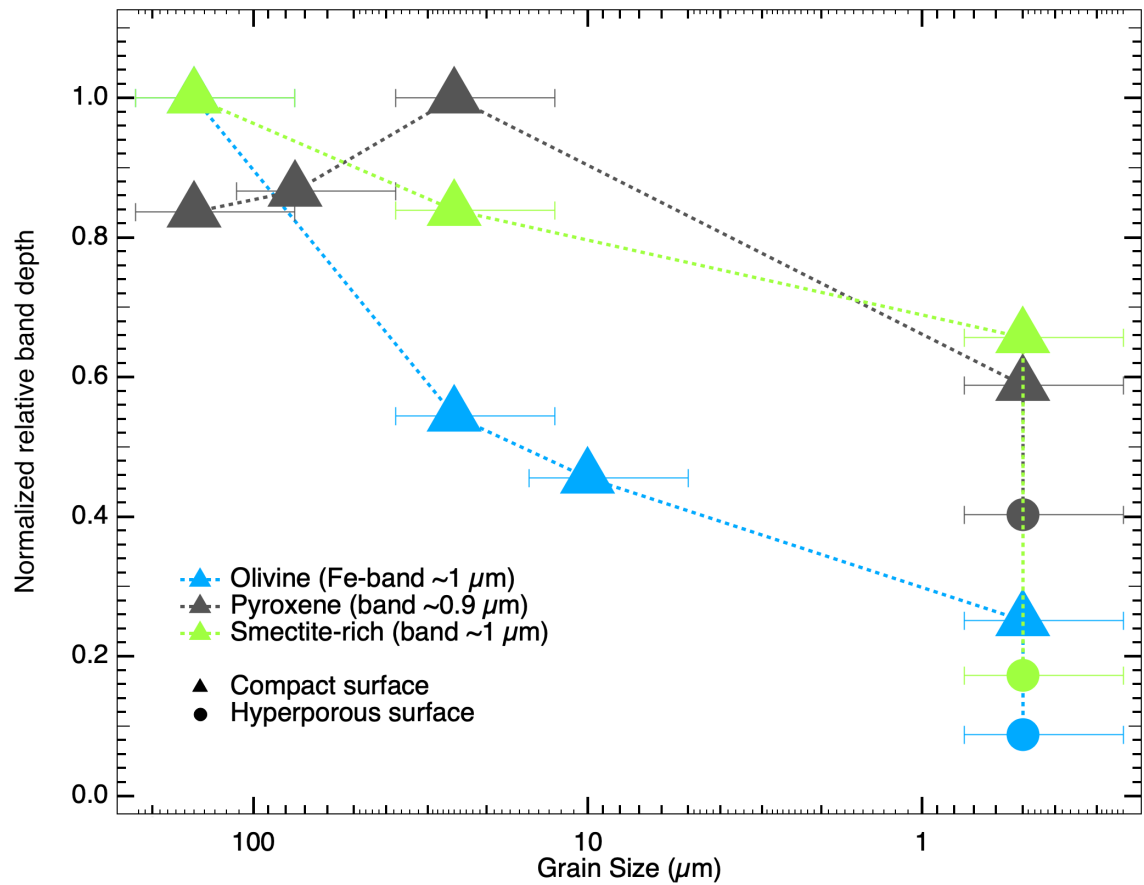

**Figure 10 :** Normalized relative band depth for the materials with different grain sizes in a compact surface (triangles) as well as with hyperfine grains in a hyperporous surface (circles). All materials have a similar behaviour, the relative band depth decreasing with grain size. When the porosity increases (i.e. hyperporous surface) the bands are even shallower. Values are presented in Table 2.

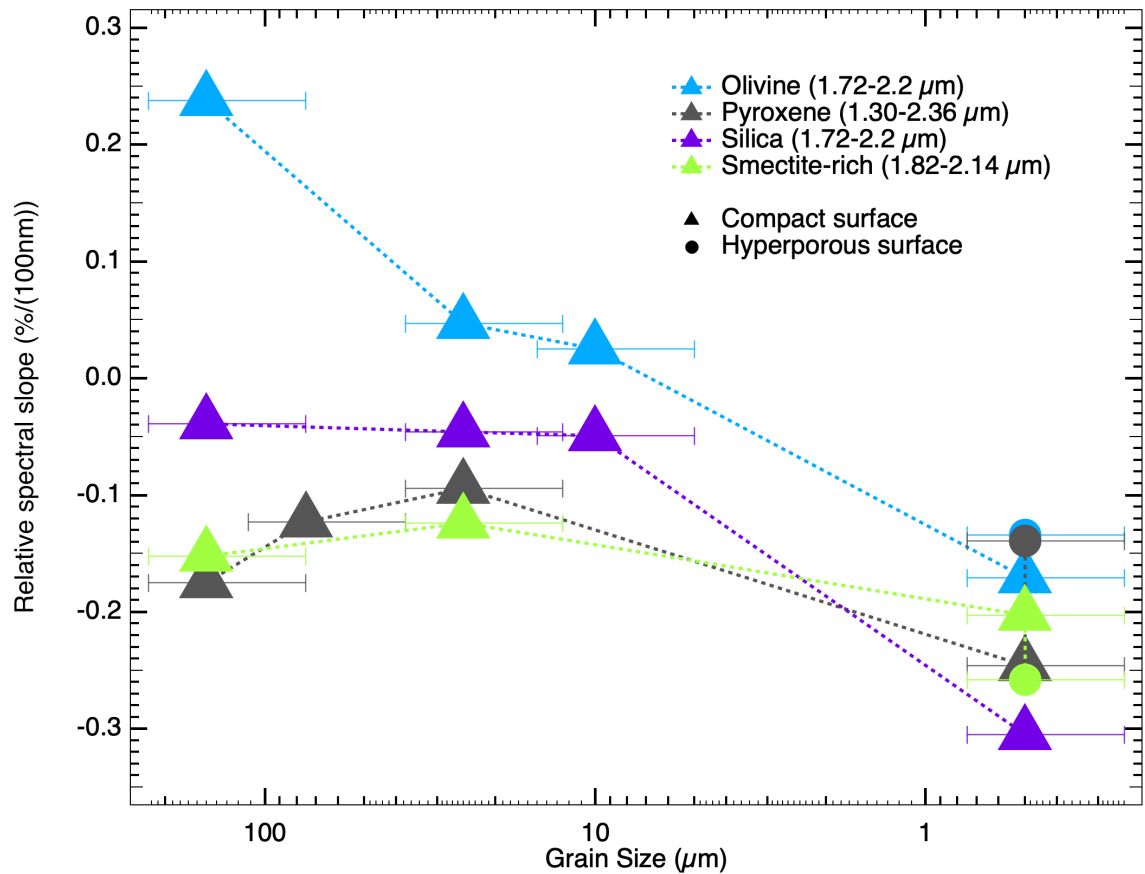

**Figure 11 :** Relative spectral slopes for the materials with different grain sizes in a compact surface (triangles) as well as with hyperfine grains in a hyperporous surface (circles). Relative spectral slopes present a similar trend for every material when the grain size is decreasing. The spectra become bluer for sub-micrometre-sized grains with respect to the samples constituted of larger grains. However, when the porosity increases, we do not observe a blueing effect with respect to the compact surface of sub-micrometre-sized grains, the slopes are very similar or even redder.

# 4. Discussion

## 4.1. Effect of grain size

As described in the previous section, grain size has a significant influence on the reflectance spectra. With decreasing grain size, we observe three effects:

- an increase of the reflectance in the visible and in the very near infrared (0.6-0.8 µm),

- a decrease of the band contrast (band depths are much shallower),

- a change of spectral slope, which becomes bluer (toward negative values).

When grain size is decreased, at constant porosity, the volume density of grains (the number of grains for a given volume) is increased, and so does the volume density of interfaces (facets of grains) on which light can be reflected. In a geometric optics regime, the probability of a photon to be scattered back to the detector is a competition between probability of scattering and probability of absorption by grains. When grains become smaller, light will be less likely to travel long distances inside grains and then be absorbed. As a consequence, reflectance is generally higher for smaller grains and band depths are smaller, as observed in our experiments for non-hyperfine grains. In the case of the hyperfine grains, the situation is somehow different.

In the case of the sub-µm sized samples, the decrease of band depth continues when compared to larger grains, even if the regime of light-scattering is expected to be different (i.e. no more in the geometric optics regime). The reflectance in the visible is higher or of the order of the reflectance of the larger grains. However, when considering the near infrared wavelengths, the reflectance appears to decrease when compared to the larger grain sizes. This reveals the presence of a spectral effect, a blueing, related to the presence of hyperfine grains.

The blueing of the spectra may be explained by the fact that with grain sizes smaller than the wavelength, absorption and scattering are not decreasing at the same rate; for a given grain size the scattering efficiency scales with $\lambda^{-4}$ (Hapke, 1993; Mustard and Hays, 1997; Strutt, 1871) while the absorption efficiency decreases with $\lambda^{-1}$ (Hapke, 1993; Mustard and Hays, 1997). If both absorption and scattering efficiencies were decreasing at the same rate, light would travel deeper in the particulate surface but would still be able to "escape" back as efficiently to the detector. However, since absorption efficiency does not decrease as fast as scattering efficiency with wavelength, light penetrates deeper in the sample, is absorbed by the particles, which should result in a net decrease of the reflectance with increasing wavelength for a given grain size, so introducing a blue slope in the spectrum. This effect was observed in Mustard and Hays, (1997) in the mid-infrared for 0-5 µm olivine grains and is observed here in the Vis-NIR for our 200-400 nm olivine grains. However, the magnitude of the observed blueing effect is modest while wavelength changes by a factor of about 10 across the 0.4-4.2 µm spectral range of our measurements. This suggests that we are probably not in a pure Rayleigh scattering regime. One possible explanation is that grains do not interact with light individually but rather collectively and that the typical scale with which a photon interact if that of the grain aggregate that scatters light coherently (Salisbury and Wald, 1992). This hypothesis can be tested by separating individual grains, which can be done by trying to increase porosity in the sample as discussed in the following section.

## 4.2. Effects of porosity

The spectra of the hyperfine and hyperporous residues show shallower absorption bands (**Figure 9**) than hyperfine compact powders. As an example, the band depth of olivine is decreased by about a factor of three between hyperfine compact powder and hyperfine and hyperporous residue.

When increasing the porosity in the sample, we increase on average the space separating grains (we increase both the macro- and the micro-porosities). By doing so, we decrease the amount of large aggregates (of hundreds of micrometres wide containing many grains), and we produce more of the small aggregates a few micrometres large composed of a few grains.

This may lead to a reduction in the individual size of the light scatterers towards that of the aggregates (a few micrometres) or to that of the grains (hundreds of nanometres), hence amplifying the effect of the decreasing grain size as we observed on the spectra of sublimation residues (**Figure 9**). Such observations were already made by Salisbury and Wald, (1992) and Cooper and Mustard, (1999). They were explained as a change in the scattering regime; small grains scatter as individual grains when separated by a distance comparable or larger than the wavelength as in a porous surface. In addition, the size of aggregates is reduced when porosity increases, decreasing their equivalent scattering size.

Interestingly, the behaviour observed is not that of a purely Rayleigh scattering regime, in which a reduction of band depth would also be expected to be associated to a strong blueing of the spectra as well as a strong decrease of reflectance. This is likely related to the fact that the size parameter $X = \frac{2\pi D}{\lambda}$ is not $\ll 1$ and that we are rather in the resonant regime discussed by Hapke, (1993) which is at the transition between Rayleigh regime and geometrics optics.

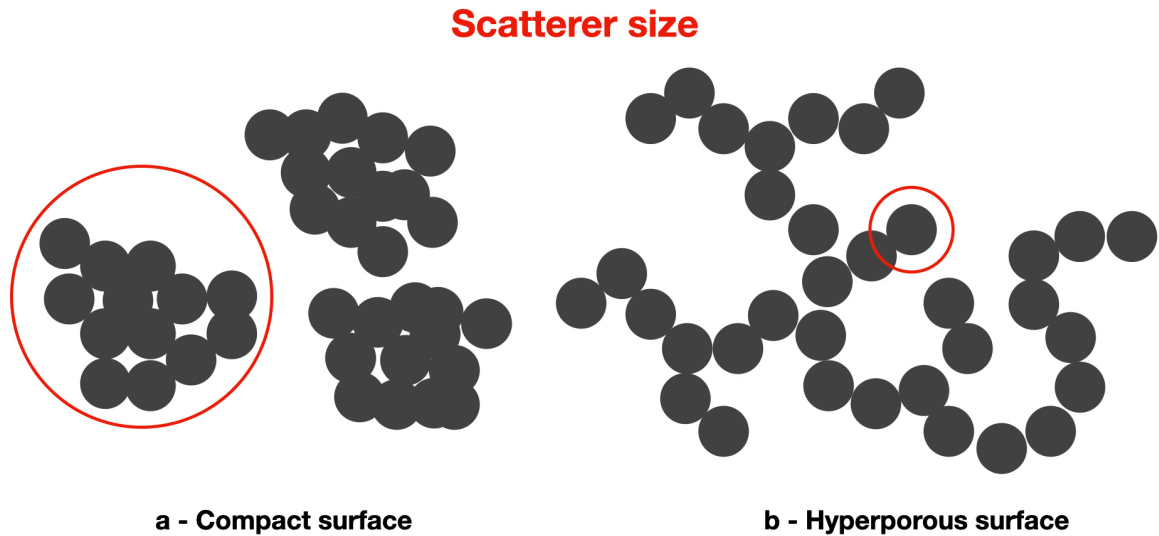

**Figure 12 :** Diagram explaining the decrease of the scatterer size down to small aggregates or to individual grains when the porosity increases. **(a)** In a regular powder (compact surface) large compact aggregates containing many grains will scatter light as if they were a single grain of equivalent size. **(b)** In a sublimation residue (hyperporous surface), the grains are more spaced, and the light will be scattered by smaller aggregates (constituted of a few grains) or by individual grains, amplifying the effects of light scattering by small grains on the reflectance spectra.

## 4.3 A Mie-theory based model

In order to have a more quantitative understanding of the behaviour observed in our experiments, we performed numerical modelling in the framework of Mie theory, using a Mie code developed by Sumlin et al. (2018). Mie theory enables to describe light scattering by isolated spherical particles in a wide range of grain sizes. We computed the single scattering albedo (**SSA**) and the phase function (**PF**) for grains with a size ranging from 0.1 to 5 µm, using the complex refractive index of a San-Carlos olivine from Zeidler et al. (2011), over the 0.5 to 2.5 µm spectral range.

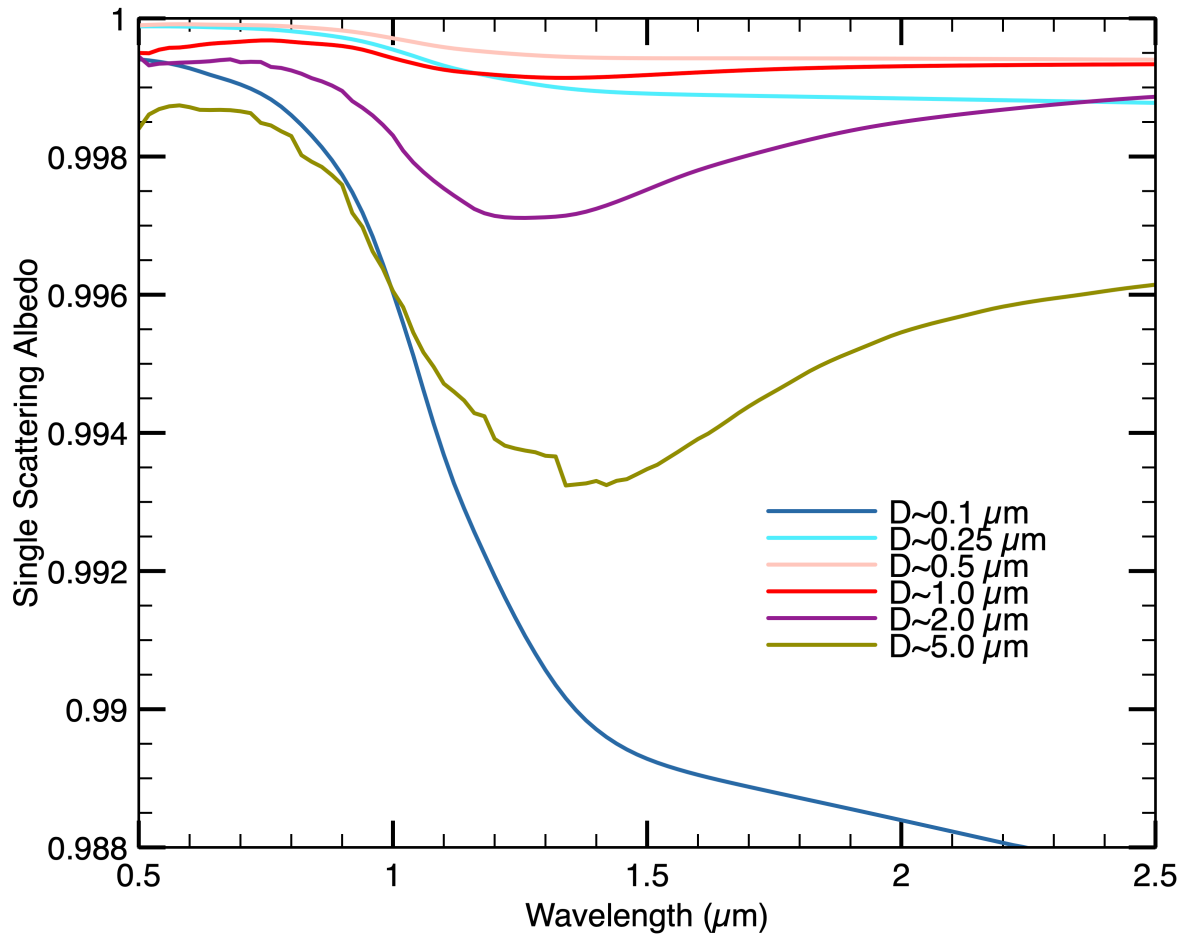

**Figure 13 :** Variation of the single scattering albedo of olivine grains with wavelength. When the grain diameter (D) decreases and starts to be of the order of the wavelength (D = 0.5-1 µm, size parameter X ≤ 5) the SSA becomes featureless and close to 1 over all the Vis-NIR spectral range (D = 0.5 – 1 µm). When the size becomes smaller than the wavelength (D = 0.1 µm), the SSA shows a strong blue slope in the NIR part of the spectrum.

With decreasing grain size, the value of SSA first increases across the whole spectral range until the size parameter $X = \frac{2\pi D}{\lambda}$ approaches 1 (D = 0.5 - 1 µm). As the SSA increases, the depth of the olivine $Fe^{2+}$ feature progressively decreases, until the absorption band almost disappears. Then, for grain sizes below 0.5 µm a strong blueing and decrease of reflectance in the high wavelength range is observed.

A second important aspect is that the single particle phase function will also change with the size-parameter, which should induce an additional spectral effect on bi-directional reflectance.

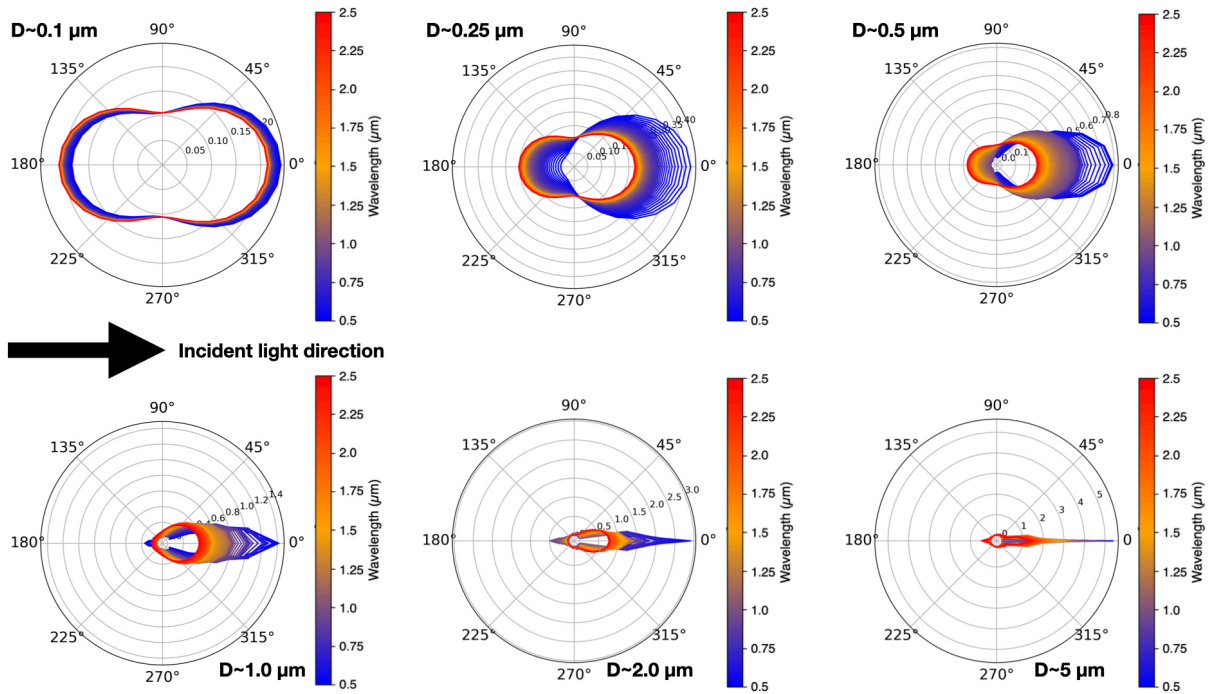

**Figure 14 :** Normalized single particle phase function computed for a single isolated grain of size *D* over the whole olivine spectrum from 0.5µm (blue) to 2.5 µm (red). Incident light comes from the left. With the decrease in the grain size, the phase function is less forward scattering and more symmetrical, close to a Rayleigh scattering phase function.

As shown in **Figure 14**, the single particle phase function presents a strong dependency with both grain size and wavelength. For grains large compared to the wavelength (X >> 1), Mie computations predict a strongly forward phase function, because of the diffraction peak. But with decreasing grain size, scattering becomes less forward and tends to be symmetrical when the scatterer size approaches that of the Rayleigh scattering regime (X << 1). In the Rayleigh regime (D < 0.1 μm), the single particle phase function is also independent of grain size.

For a given grain sizes in the 0.5-5 μm range, the single particle phase function strongly depends on the wavelength. As an example, in the case of the 0.5 μm grain diameter, the scattering behaviour is expected to change from forward to symmetrical as wavelength increases from 0.5 to 2.5 μm. When observed in reflectance, this effect should produce a reddening of the spectra.

This Mie modelling enables to show that the spectral behaviour of an isolated grain in the resonant size parameter range will be the competition of two phenomena: the decrease of scattering efficiency with increasing wavelength, and the change in the shape of the single particle phase function. Understanding how these two effects influence the bi-directional reflectance spectra requires more modelling effort, with the simulation of the behaviour of a collection of grains. This will be the subject of a future work.

## 5. Implications for cometary and asteroidal observations

We have previously shown that the presence of hyperfine scatterers drastically reduces the band depths and modifies the spectral slope of all silicates studied here and that increasing the space between the grains in the surface amplifies these effects.

When interpreting spectral observations in the Vis-NIR, we must consider these effects of grain size and porosity.

## 5.1 The D-type signature cannot be explained by grain size alone

One of the possible explanations proposed for the low albedo of comets is the presence of a surface composed of hyperfine grains, of sizes below visible wavelength (Greenberg, 1986). Mustard and Hays (1997) already observed this effect for grains finer than 5 μm, although reflectance as low as a few percent were not observed. The reflectance spectra in the mid-infrared of the hyperfine samples studied here, having grain sizes of the order of 200-400 nm, show a similar behaviour, even more pronounced (in a future paper currently under preparation). However, no darkening was observed in the visible for these hyperfine powders. On the contrary, we noticed an increase of the reflectance between 0.6 and 1.2 μm, due to the shrinking width of the absorption features and the larger amount of grain interfaces. The presence of porosity does not help in darkening the sample as shown by our results on sublimation residues. Therefore, an absorbing component is needed in the cometary material to explain the dark nature of cometary dust. Large cometary dust grains ($100 - 600$ μm) observed by the COSIMA/Cosiscope were found to be absorbing and porous with a mean free path in the $20 - 25$ μm range (Langevin et al., 2017). This absorbing constituent in the visible range can be carbon-rich compounds, but the presence of small Fe-rich opaques may probably also play a role in order to explain the darkness of the nuclei up to 4 μm at least (Beck et al., 2018; Quirico et al., 2016; Rousseau et al., 2018).

## 5.2. The small grain spectral degeneracy

The results obtained on hyperfine grains all reveal that absorption bands present in large-grained samples tend to diminish strongly when grain size decreases. This has been shown previously for sieved fractions of various minerals, and the present study shows that this effect continues when considering hyperfine particles. One of the most striking examples is the pyroxene sample (**Figure 7**b). For this sample, two deep bands (20-30% depth) with similar depth can be observed at 1 and 2 μm. However, in the case of the hyperfine-grained sublimation residue the 1 μm band depth is reduced to only 9% while the 2 μm band depth is barely detectable. In the case of olivine, while the 1 μm band depth is almost 80% for the 100−200 μm size fraction, the depth is reduced down to 7% for the sublimation residue made of sub-micrometre-sized grains (**Figure 9 & Figure 10**). These absorptions are due to $Fe^{2+}$ transitions in the silicate crystal field and have values of the extinction coefficient $k$ typically in the range $10^{-5} − 10^{-3}$ for these silicate compositions. This means that for hyperfine grains of compounds with k values in this range, it is expected that any absorption feature will be difficult to detect from reflectance spectra. One compound of special interest with k values in this range is water-ice in the Vis-NIR (Grundy and Schmitt, 1998). Although accurate radiative transfer modelling would be required to assess more firmly this point, our results suggest that reflectance spectra of a particulate surface of hyperfine ice grains (< 200 nm) would show very faint absorptions in the Vis-NIR, if any. Water "ash deposit" from plumes operating at the surface of icy satellites may be difficult to detect for that reason. The case where $k > 0.03$ (i.e. strongly absorbing material) was not investigated in our work except for the smectite-rich sample in the 3 μm region (**Figure 15**). In that case, the presence of a significant fraction of phyllosilicate (smectite) explains the presence of a 2.7 μm fundamental OH-stretching absorption. This feature is almost saturated in the large grains spectra we measured (reflectance of 0.01 at the bottom of the band), which is typically what is observed for the OH-feature in phyllosilicates. In the case of its sublimation residue, this band appears significantly desaturated (reflectance of 0.07 at the bottom of the band), but the band remains very strong (relative depth of 68 %) as seen in **Figure 15**.

If a material is in the form of hyperfine grains in a hyperporous surface, visible and near infrared reflectance spectra from this surface may not show any hint of its presence. Consequently, the absence of the absorption bands of this material on the reflectance spectra of this surface does not imply the absence of this material. Moreover, in the case where absorption bands are observed, the quantification of the material causing these bands could also be strongly complicated by the influences of grain size and porosity on the band depths.

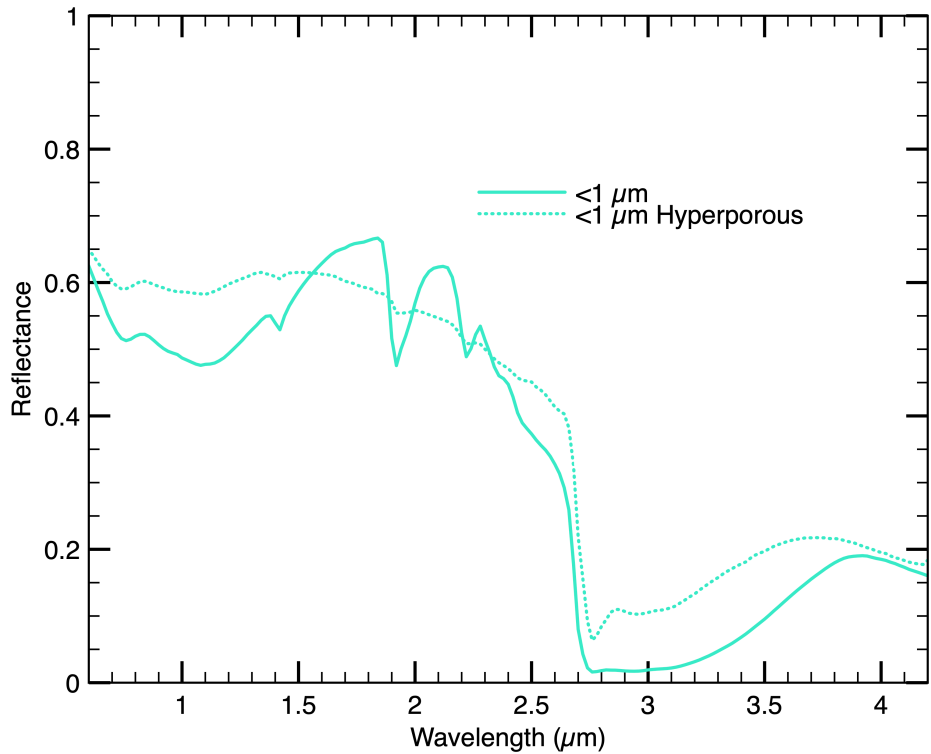

**Figure 15 :** Reflectance spectra of two surfaces made of the smectite-rich material: a regular (compact) surface of hyperfine grains (solid line) and a sublimation residue of hyperfine grains (dotted line). Around 2.7 µm the phyllosilicate OH absorption feature is significantly reduced in the sublimation residue with respect to the compact surface.

## 5.3. Can small grains explain the spectra of B-types?

B-type asteroids are a class of dark objects belonging to the C-complex. Their reflectance spectra are reminiscent of that of C-type however with an overall blue spectral slope and the lack of phyllosilicate related absorptions at 0.7, 0.9 and 1.1 µm. Asteroids belonging to the B-type may present phyllosilicates at their surface based on the presence of a 3-µm band, as in the case of Pallas or Bennu (Hamilton et al., 2019). Interestingly, the spectra obtained on our phyllosilicate-rich sublimation residue (the smectite-rich material) shows a blueing together with a decrease of the 0.7 − 0.9 and 1.1 μm absorption features, while still preserving the OH-band. These results may suggest that the surface of some B-type asteroids may be constituted of hyperporous and hyperfine grained materials. This would imply that such surfaces may be loosely consolidated and would not deliver meteorites but rather IDPs (Vernazza and Beck, 2016). The B-type asteroid most studied to date, Bennu, however does not show strong evidence for the presence of fine grains, but rather large rock fragments with sizes around tens of centimetres to meters (DellaGiustina et al., 2019). Unpowdered meteorite samples (i.e. rock) tend to show blue slopes which may explain the spectral signature of Bennu (DellaGiustina et al., 2019). Because Bennu is a Near-Earth asteroid (NEA), geological processes specific to NEAs may explain a rock-dominated surface (collisional breakup and re-assembly, surface refreshing by close encounters, dust migration due to rotation, etc.). In the case of main-belt asteroids, the surface is expected to be more particulate and a rock-dominated surface is more improbable. In that case, the presence of a hyperfine and hyperporous surface, possibly produced by volatile sublimation or cryo-volcanism, constitutes an interesting alternative to explain the blue slope of B-type.

## 5.4. The surface texture of S- and V-type asteroids

S-type asteroids are a group of undifferentiated asteroids. They are believed to be the main parent bodies reservoir of ordinary chondrites. Reflectance spectra in the Vis-NIR of S-type asteroids are relatively bright (around 10 %), present deep absorption bands around 1 and 2.2 μm that can be related to the presence of silicates, and display a strong red spectral slope (DeMeo et al., 2009). V-type are expected to be basalt-covered asteroids, showing strong surface signatures of ortho-pyroxene based on absorption at 0.9 and 2 µm. Vis-NIR spectroscopic measurements on powders constituted of sub-micrometre-sized grains of silicates (both olivine and pyroxene) show very shallow bands around 1 and 2.2 µm, down to 7 % for the Fe-band of the olivine. Spectral slopes of these samples also tend toward negative values (blue slope) which are not compatible with reflectance spectra of asteroids of the S-type, since they present a strong red slope. Collisions were the most important geological process occurring on these surfaces. When a hypervelocity collision occurs, the target experiences a significant comminution and the fragments are expected to have a power-law grain size distribution. The smallest fragments are the most numerous and they can have sizes below 1-µm as revealed by cratering experiments (Buhl et al., 2014). Following that line, one may expect such surfaces to be covered by sub-µm grains, which, according to our experiments, should not display strong spectral signatures. Our results imply that the surfaces of most of S-type asteroids are not constituted of hyperfine grains, but these fine grains may be close-packed together and thus behave optically like larger grains (**Error! Reference source not found.**). This later hypothesis is appealing since impact will also tend to compact the surface material and aggregate finer grains to each other.

# 6. Conclusion

In this article we attempted to synthesize surfaces presenting similarities to what is and might be observed on comets and C-, P- and D-type asteroids, in terms of texture, micro-structures and spectral characteristics, and to determine precisely the effects of grain size and surface porosity on Vis-NIR reflectance spectra.

We first elaborated an experimental protocol to produce powders made of grains with an individual size smaller than the micrometre. Analogues of cometary dust micro-structures have been made by increasing the porosity of these powders by sublimation of "dirty" water ice particles, a method that mimics the process occurring on cometary surfaces.

Spectroscopic measurements on powders of decreasing grain size from 200 µm to 25 µm have shown that grain size strongly influences the shape of reflectance spectra by reducing the depth of the absorption features and changing the spectral slope toward negative values (spectral blueing). When the grain size decreases even more and becomes smaller than about one micrometre, these changes continue and become even more marked. We have also shown that porosity amplifies the decrease of band contrast so much that the weakest absorption features tend to disappear.

We explain these spectral changes by a change in scattering regime due to the reduction in size of the scatterers when grains become hyperfine. In addition, the amplification of these effects when the surface porosity increases may come from a further reduction of the size of individual scatterers, from large (tens to hundreds of micrometres) and compact aggregates of tens to hundreds of grains in the compact surfaces, to tiny (some micrometres) and porous aggregates of only few sub-micrometre-sized grains in the hyperporous sublimation residues.

This work highlights that studies of composition from spectroscopic observations on small bodies might not be as simple as a search for spectral signatures (absorption bands of components). Indeed, sub-micrometre grain size and surface porosity can hide the spectral signatures of some components since they tend to make most of silicate spectra relatively blue and featureless. Observers must therefore keep in mind that the absence of the absorption band(s) of a component from a spectrum of a small body, does not necessarily means that this component is absent from its surface. The effects highlighted in this paper can also provide insights to better understand the texture, the activity and the formation scenarios of small bodies of the Solar System from the analysis of their spectroscopic observations.

Future experimentations will explore these grain size and hyperporosity effects in mixtures of several relevant endmembers containing dark materials (opaque minerals, organic matter) as well as in even more porous samples to explore light scattering by individual grains, using the non-absorbing optical properties of potassium bromide (KBr).

## Acknowledgements

This work was funded by the European Research Council under the H2020 framework program/ERC grant agreement no. 771691 (Solarys). Additional support by the Programme National de Planétologie and the Centre National d'Etude Spatiale is acknowledged. This manuscript greatly beneficiated from comments by two reviewers. A special thanks to H. Capelo and C. Cesar from Univesität Bern for providing pyroxene.

## Data Availability

The data of all the measured and modelled reflectance spectra shown in this article can be found on the Grenoble Astrophysics and Planetology Solid Spectroscopy and Thermodynamics (GhoSST) database hosted in the Solid Spectroscopy Hosting Architecture of Databases and Expertise (SSHADE) at in Sultana, Robin (2018): Vis-NIR reflectance spectra of powdered olivine, pyroxene, smectite and silica at 4 different grain sizes (from sub-µm to 200 µm) and with sub-µm grains in compact or porous surfaces.

SSHADE/GhoSST (OSUG Data Center).Dataset/SpectralData.

https://doi.org/10.26302/SSHADE/EXPERIMENT_OP_20200908_001.

DellaGiustina, D.N., Emery, J.P., Golish, D.R., Rozitis, B., Bennett, C.A., Burke, K.N., Ballouz, R.L., Becker, K.J., Christensen, P.R., Drouet d'Aubigny, C.Y., Hamilton, V.E., Reuter, D.C., Rizk, B., Simon, A.A., Asphaug, E., Bandfield, J.L., Barnouin, O.S., Barucci, M.A., Bierhaus, E.B., Binzel, R.P., Bottke, W.F., Bowles, N.E., Campins, H., Clark, B.C., Clark, B.E., Connolly, H.C., Daly, M.G., Leon, J. de, Delbo', M., Deshapriya, J.D.P., Elder, C.M., Fornasier, S., Hergenrother, C.W., Howell, E.S., Jawin, E.R., Kaplan, H.H., Kareta, T.R., Le Corre, L., Li, J.Y., Licandro, J., Lim, L.F., Michel, P., Molaro, J., Nolan, M.C., Pajola, M., Popescu, M., Garcia, J.L.R., Ryan, A., Schwartz, S.R., Shultz, N., Siegler, M.A., Smith, P.H., Tatsumi, E., Thomas, C.A., Walsh, K.J., Wolner, C.W.V., Zou, X.D., Lauretta, D.S., Highsmith, D.E., Small, J., Vokrouhlický, D., Brown, E., Donaldson Hanna, K.L., Warren, T., Brunet, C., Chicoine, R.A., Desjardins, S., Gaudreau, D., Haltigin, T., Millington-Veloza, S., Rubi, A., Aponte, J., Gorius, N., Lunsford, A., Allen, B., Grindlay, J., Guevel, D., Hoak, D., Hong, J., Schrader, D.L., Bayron, J., Golubov, O., Sánchez, P., Stromberg, J., Hirabayashi, M., Hartzell, C.M., Oliver, S., Rascon, M., Harch, A., Joseph, J., Squyres, S., Richardson, D., McGraw, L., Ghent, R., Asad, M.M.A., Johnson, C.L., Philpott, L., Susorney, H.C.M., Cloutis, E.A., Hanna, R.D., Ciceri, F., Hildebrand, A.R., Ibrahim, E.M., Breitenfeld, L., Glotch, T., Rogers, A.D., Ferrone, S., Fernandez, Y., Chang, W., Cheuvront, A., Trang, D., Tachibana, S., Yurimoto, H., Brucato, J.R., Poggiali, G., Dotto, E., Epifani, E.M., Crombie, M.K., Lantz, C., Izawa, M.R.M., de Leon, J., Clemett, S., Thomas-Keprta, K., Van wal, S., Yoshikawa, M., Bellerose, J., Bhaskaran, S., Boyles, C., Chesley, S.R., Farnocchia, D., Harbison, A., Kennedy, B., Knight, A., Martinez-Vlasoff, N., Mastrodemos, N., McElrath, T., Owen, W., Park, R., Rush, B., Swanson, L., Takahashi, Y., Velez, D., Yetter, K., Thayer, C., Adam, C., Antreasian, P., Bauman, J., Bryan, C., Carcich, B., Corvin, M., Geeraert, J., Hoffman, J., Leonard, J.M., Lessac-Chenen, E., Levine, A., McAdams, J.,

McCarthy, L., Nelson, D., Page, B., Pelgrift, J., Sahr, E., Stakkestad, K., Stanbridge, D., Wibben, D., Williams, B., Williams, K., Wolff, P., Hayne, P., Kubitschek, D., Fulchignoni, M., Hasselmann, P., Merlin, F., Praet, A., Billett, O., Boggs, A., Buck, B., Carlson-Kelly, S., Cerna, J., Chaffin, K., Church, E., Coltrin, M., Daly, J., Deguzman, A., Dubisher, R., Eckart, D., Ellis, D., Falkenstern, P., Fisher, A., Fisher, M.E., Fleming, P., Fortney, K., Francis, S., Freund, S., Gonzales, S., Haas, P., Hasten, A., Hauf, D., Hilbert, A., Howell, D., Jaen, F., Jayakody, N., Jenkins, M., Johnson, K., Lefevre, M., Ma, H., Mario, C., Martin, K., May, C., McGee, M., Miller, B., Miller, C., Miller, G., Mirfakhrai, A., Muhle, E., Norman, C., Olds, R., Parish, C., Ryle, M., Schmitzer, M., Sherman, P., Skeen, M., Susak, M., Sutter, B., Tran, Q., Welch, C., Witherspoon, R., Wood, J., Zareski, J., Arvizu-Jakubicki, M., Audi, E., Bandrowski, R., Becker, T.L., Bendall, S., Bloomenthal, H., Blum, D., Boynton, W. V., Brodbeck, J., Chojnacki, M., Colpo, A., Contreras, J., Cutts, J., Dean, D., Diallo, B., Drinnon, D., Drozd, K., Enos, H.L., Enos, R., Fellows, C., Ferro, T., Fisher, M.R., Fitzgibbon, G., Fitzgibbon, M., Forelli, J., Forrester, T., Galinsky, I., Garcia, R., Gardner, A., Habib, N., Hamara, D., Hammond, D., Hanley, K., Harshman, K., Herzog, K., Hill, D., Hoekenga, C., Hooven, S., Huettner, E., Janakus, A., Jones, J., Kidd, J., Kingsbury, K., Balram-Knutson, S.S., Koelbel, L., Kreiner, J., Lambert, D., Lewin, C., Lovelace, B., Loveridge, M., Lujan, M., Maleszewski, C.K., Malhotra, R., Marchese, K., McDonough, E., Mogk, N., Morrison, V., Morton, E., Munoz, R., Nelson, J., Padilla, J., Pennington, R., Polit, A., Ramos, N., Reddy, V., Riehl, M., Roper, H.L., Salazar, S., Selznick, S., Stewart, S., Sutton, S., Swindle, T., Tang, Y.H., Westermann, M., Worden, D., Zega, T., Zeszut, Z., Bjurstrom, A., Bloomquist, L., Dickinson, C., Keates, E., Liang, J., Nifo, V., Taylor, A., Teti, F., Caplinger, M., Bowles, H., Carter, S., Dickenshied, S., Doerres, D., Fisher, T., Hagee, W., Hill, J., Miner, M., Noss, D., Piacentine, N., Smith, M., Toland, A., Wren, P.,

Ghent, R., Binzel, R.P., Asad, M.M.A., Johnson, C.L., Philpott, L., Susorney, H.C.M., Cloutis, E.A., Hanna, R.D., Connolly, H.C., Ciceri, F., Hildebrand, A.R., Ibrahim, E.M., Breitenfeld, L., Glotch, T., Rogers, A.D., Clark, B.E., Ferrone, S., Thomas, C.A., Campins, H., Fernandez, Y., Chang, W., Cheuvront, A., Trang, D., Tachibana, S., Yurimoto, H., Brucato, J.R., Poggiali, G., Pajola, M., Dotto, E., Mazzotta Epifani, E., Crombie, M.K., Lantz, C., Izawa, M.R.M., de Leon, J., Licandro, J., Garcia, J.L.R., Clemett, S., Thomas-Keprta, K., Van wal, S., Yoshikawa, M., Bellerose, J., Bhaskaran, S., Boyles, C., Chesley, S.R., Elder, C.M., Farnocchia, D., Harbison, A., Kennedy, B., Knight, A., Martinez-Vlasoff, N., Mastrodemos, N., McElrath, T., Owen, W., Park, R., Rush, B., Swanson, L., Takahashi, Y., Velez, D., Yetter, K., Thayer, C., Adam, C., Antreasian, P., Bauman, J., Bryan, C., Carcich, B., Corvin, M., Geeraert, J., Hoffman, J., Leonard, J.M., Lessac-Chenen, E., Levine, A., McAdams, J., McCarthy, L., Nelson, D., Page, B., Pelgrift, J., Sahr, E., Stakkestad, K., Stanbridge, D., Wibben, D., Williams, B., Williams, K., Wolff, P., Hayne, P., Kubitschek, D., Deshapriya, J.D.P., Fornasier, S., Fulchignoni, M., Hasselmann, P., Merlin, F., Praet, A., Bierhaus, E.B., Billett, O., Boggs, A., Buck, B., Carlson-Kelly, S., Cerna, J., Chaffin, K., Church, E., Coltrin, M., Daly, J., Deguzman, A., Dubisher, R., Eckart, D., Ellis, D., Falkenstern, P., Fisher, A., Fisher, M.E., Fleming, P., Fortney, K., Francis, S., Freund, S., Gonzales, S., Haas, P., Hasten, A., Hauf, D., Hilbert, A., Howell, D., Jaen, F., Jayakody, N., Jenkins, M., Johnson, K., Lefevre, M., Ma, H., Mario, C., Martin, K., May, C., McGee, M., Miller, B., Miller, C., Miller, G., Mirfakhrai, A., Muhle, E., Norman, C., Olds, R., Parish, C., Ryle, M., Schmitzer, M., Sherman, P., Skeen, M., Susak, M., Sutter, B., Tran, Q., Welch, C., Witherspoon, R., Wood, J., Zareski, J., Arvizu-Jakubicki, M., Asphaug, E., Audi, E., Ballouz, R.L., Bandrowski, R., Becker, K.J., Becker, T.L., Bendall, S., Bennett, C.A., Bloomenthal, H., Blum, D., Boynton, W. V., Brodbeck, J., Burke, K.N., Chojnacki,

M., Colpo, A., Contreras, J., Cutts, J., Drouet d'Aubigny, C.Y., Dean, D., DellaGiustina, D.N., Diallo, B., Drinnon, D., Drozd, K., Enos, H.L., Enos, R., Fellows, C., Ferro, T., Fisher, M.R., Fitzgibbon, G., Fitzgibbon, M., Forelli, J., Forrester, T., Galinsky, I., Garcia, R., Gardner, A., Golish, D.R., Habib, N., Hamara, D., Hammond, D., Hanley, K., Harshman, K., Hergenrother, C.W., Herzog, K., Hill, D., Hoekenga, C., Hooven, S., Howell, E.S., Huettner, E., Janakus, A., Jones, J., Kareta, T.R., Kidd, J., Kingsbury, K., Balram-Knutson, S.S., Koelbel, L., Kreiner, J., Lambert, D., Lauretta, D.S., Lewin, C., Lovelace, B., Loveridge, M., Lujan, M., Maleszewski, C.K., Malhotra, R., Marchese, K., McDonough, E., Mogk, N., Morrison, V., Morton, E., Munoz, R., Nelson, J., Nolan, M.C., Padilla, J., Pennington, R., Polit, A., Ramos, N., Reddy, V., Riehl, M., Rizk, B., Roper, H.L., Salazar, S., Schwartz, S.R., Selznick, S., Shultz, N., Smith, P.H., Stewart, S., Sutton, S., Swindle, T., Tang, Y.H., Westermann, M., Wolner, C.W.V., Worden, D., Zega, T., Zeszut, Z., Bjurstrom, A., Bloomquist, L., Dickinson, C., Keates, E., Liang, J., Nifo, V., Taylor, A., Teti, F., Caplinger, M., Bowles, H., Carter, S., Dickenshied, S., Doerres, D., Fisher, T., Hagee, W., Hill, J., Miner, M., Noss, D., Piacentine, N., Smith, M., Toland, A., Wren, P., Bernacki, M., Munoz, D.P., Watanabe, S.I., Sandford, S.A., Aqueche, A., Ashman, B., Barker, M., Bartels, A., Berry, K., Bos, B., Burns, R., Calloway, A., Carpenter, R., Castro, N., Cosentino, R., Donaldson, J., Dworkin, J.P., Elsila Cook, J., Emr, C., Everett, D., Fennell, D., Fleshman, K., Folta, D., Gallagher, D., Garvin, J., Getzandanner, K., Glavin, D., Hull, S., Hyde, K., Ido, H., Ingegneri, A., Jones, N., Kaotira, P., Lim, L.F., Liounis, A., Lorentson, C., Lorenz, D., Lyzhoft, J., Mazarico, E.M., Mink, R., Moore, W., Moreau, M., Mullen, S., Nagy, J., Neumann, G., Nuth, J., Poland, D., Reuter, D.C., Rhoads, L., Rieger, S., Rowlands, D., Sallitt, D., Scroggins, A., Shaw, G., Simon, A.A., Swenson, J., Vasudeva, P., Wasser, M., Zellar, R., Grossman, J., Johnston, G., Morris, M., Wendel, J., Burton, A., Keller, L.P.,